\documentclass[useAMS,usenatbib]{mn2e}
\usepackage{graphicx}
\title[Chemical Abundances for 18 Elements]{Parent Stars of Extrasolar Planets. VIII. Chemical Abundances for 18 Elements in 31 Stars}
\author[G. Gonzalez]{Guillermo Gonzalez$^{1}$\thanks{E-mail:
gonzog@iastate.edu},
Chris Laws$^{2}$ \\
$^{1}$Iowa State University, Department of Physics and Astronomy, Ames, 
IA 50011 \\
$^{2}$Department of Astronomy, University of Washington, Box 351580, Seattle, WA 98195}
\begin{document}

\date{Accepted ??. Received ??; in original form ??}

\pagerange{\pageref{firstpage}--\pageref{lastpage}} \pubyear{??}

\maketitle

\label{firstpage}

\begin{abstract}
We present the results of detailed spectroscopic abundance analyses for 18 elements in 31 nearby stars with planets. The resulting abundances are combined with other similar studies of nearby stars with planets and compared to a sample of nearby stars without detected planets. We find some evidence for abundance differences between these two samples for Al, Si and Ti. Some of our results are in conflict with a recent study of stars with planets in the SPOCS database. We encourage continued study of the abundance patterns of stars with planets to resolve these discrepancies.
\end{abstract}

\section{Introduction}

The present study continues our series on the chemical abundances of nearby stars with planets (SWPs); for a summary of previous papers in the series, see \citet{laws03}. To date, the only well-established chemical abundance anomaly among SWPs is the dependence of the incidence of giant planets on the host star's metallicity \citep{fv05,sant05}.

Less conclusive has been evidence for differences in the chemical abundance patterns between SWPs and stars without known planets. Several recent studies have presented the results of extensive abundance analyses of SWPs (\citet{bond06,ecu06b,gil06,gg01,luck06,rob06,takhon05}), but the results of these studies are not entirely consistent with each other. For example, \citet{rob06} reported statistically significant differences between SWPs and a comparison sample for Si/Fe and Ni/Fe abundance ratios. Other studies have reported, instead, differences for Li, Na, Mg and Al.

The observed compositional differences between SWPs and comparison stars have been discussed within the context of three classes of explanation \citep{ggpasp06}: primordial, orbital period bias and self-enrichment. The primordial explanation best accounts for the data, but the other two explanations can not yet be eliminated. Discovery of additional abundance anomalies among SWPs would help us test these three explanations more critically and allow us to determine their relative contributions. The results of these tests, in turn, will allow us to set tighter constraints on planet formation models. For example, \citet{idalin05} have reproduced the observed metallicity dependence of giant planet incidence assuming the core instability accretion model of planet formation.

In the present work we employ the stellar atmospheric parameters presented in \citet{laws03} for 31 SWPs to determine [el/H] values for 18 elements: Li, C, N, O, Na, Mg, Al, Si, S, Ca, Sc, Ti, Cr, Mn, Ni, Cu, Zn and Eu. Since most of these elements have been studied in SWPs by others, we can compare our results to published data to check for consistency. More importantly, since the lists of SWPs included in the published spectroscopic studies have considerable overlap with each other and with the present work, we can produce a new, improved database of SWP abundances by correcting for systematic abundance differences. We can do the same for published data on comparison stars. In this way, we provide more sensitive tests of the claimed abundance differences between SWPs and control samples.

\section{Abundance Analysis and Results}
\subsection{Sample and Analysis}

We focus on the 31 SWPs examined in \citet{laws03} (here, we are now including HD\,202206 amongst the SWP sample for comparison purposes in this paper). We measured equivalent widths (EWs) for atomic lines of C, N, O, Na, Mg, Al, Si, S, Ca, Sc, Ti, Cr, Ni, Zn and Eu. We then used these EWs values and the values of $T_{\rm eff}$, $\log g$, $\xi_{\rm t}$, and [Fe/H] for each star given in \citet{laws03} and  the \citet{kur93} LTE plane-parallel model atmospheres as input to the line analysis code MOOG (\citet{sneden73}, updated version) to determine the elemental abundances. This is the same code we have used in our previous papers in this series. 

Details of our method of analysis, including determination of the uncertainties of the atmospheric parameters and abundances, are described in \citet{gg98} and \citet{gv98}. In brief, for each star we determined the four basic atmospheric parameters and their uncertainties from the measured Fe I and Fe II EWs and assuming excitation and ionization equilibrium. We then propagated these uncertainties in our calculation of the uncertainties of the abundances of the other elements. The exceptions to this procedure were abundances determined from spectrum synthesis (see below); for these, we adopted an uncertainty of $\pm$ 0.10 dex, which results from the uncertainty in visually matching the observed and synthetic spectra. We applied corrections for non-LTE effects only to the measurements of the O I triplet near 7770 \AA, as prescribed by \citet{tak03}. 

We also determined [el/H] values for several elements by comparing the observed spectra with synthetic spectra, again utilizing model atmospheres and the basic stellar parameters given in \citet{laws03}. As in previous papers in this series, [Li/H] and [Al/H] were determined from syntheses of the Li region near 6707 \AA\ . For the present work, we added two more regions: $\lambda\lambda$ 5777-5787 \AA\ for the Cu abundance from the Cu I line at 5782 \AA\ and additional constraints on the Cr abundance; and $\lambda\lambda$ 6005-6015 \AA for the Mn abundance from the Mn I line at 6013 \AA\ as well as additional constraints on C and Ni abundances. Sample syntheses of these regions for HD\,82943 are shown in Figure 1.

\begin{figure}
  \includegraphics[width=3.5in]{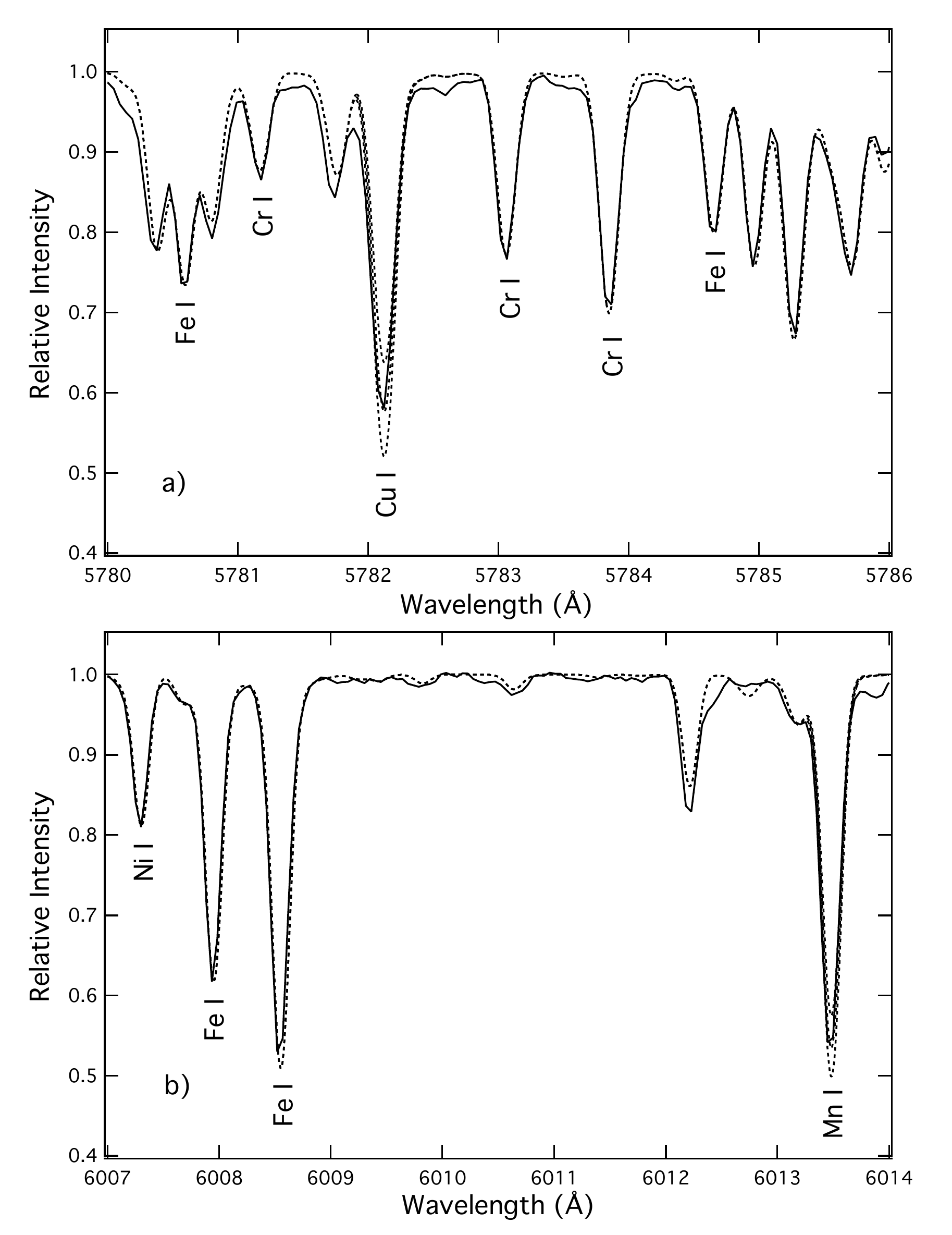}
 \caption{Sample syntheses of the spectral regions in HD\,82943 containing the Cu I (panel a) and Mn I (panel b) absorption lines used in the present work. The observed spectra are shown as solid curves, and the syntheses are shown as dotted curves. Three syntheses are shown for each element, Cu and Mn: nominal, +0.15 dex and -0.15 dex. Absorption lines of other elements analyzed in these regions are also identified.}
\end{figure}

The hyperfine components for the Cu I line are from \citet{cunha02}, and the hyperfine components for the Mn I line are from \citet{proch00}. Additional lines for these spectral intervals are from the compilations of \citet{kur95}, adjusted to provide a good match between the synthetic and observed solar spectra. Typical adjustments to the line oscillator strengths were a few tenths of a dex, but it was as large as 0.8 dex in one instance. For Mn and Cu, we adopted logarithmic solar abundances of 5.33 and 4.05, respectively. For those elements studied in previous papers in this series, we have employed the same solar abundances in order to facilitate comparison between these data sets.

\subsection{Results}

We present our abundance results in Tables 1 -- 5 and Figures 2 and 3. Where appropriate, we have listed separately in the tables the [X/H] values determined from EW analysis and those determined from spectrum synthesis. In cases where measurements from both methods are available, we calculated average values.

For some elements (O, Sc, Ti and Ni) the agreement between the present results and those of \citet{gil06} is close. For other elements (Mg, Al, Si and Ca) there are significant offsets evident in Figures 2 and 3. We correct the offsets between our results and others in the literature in the following section.

\begin{figure}
  \includegraphics[width=3.5in]{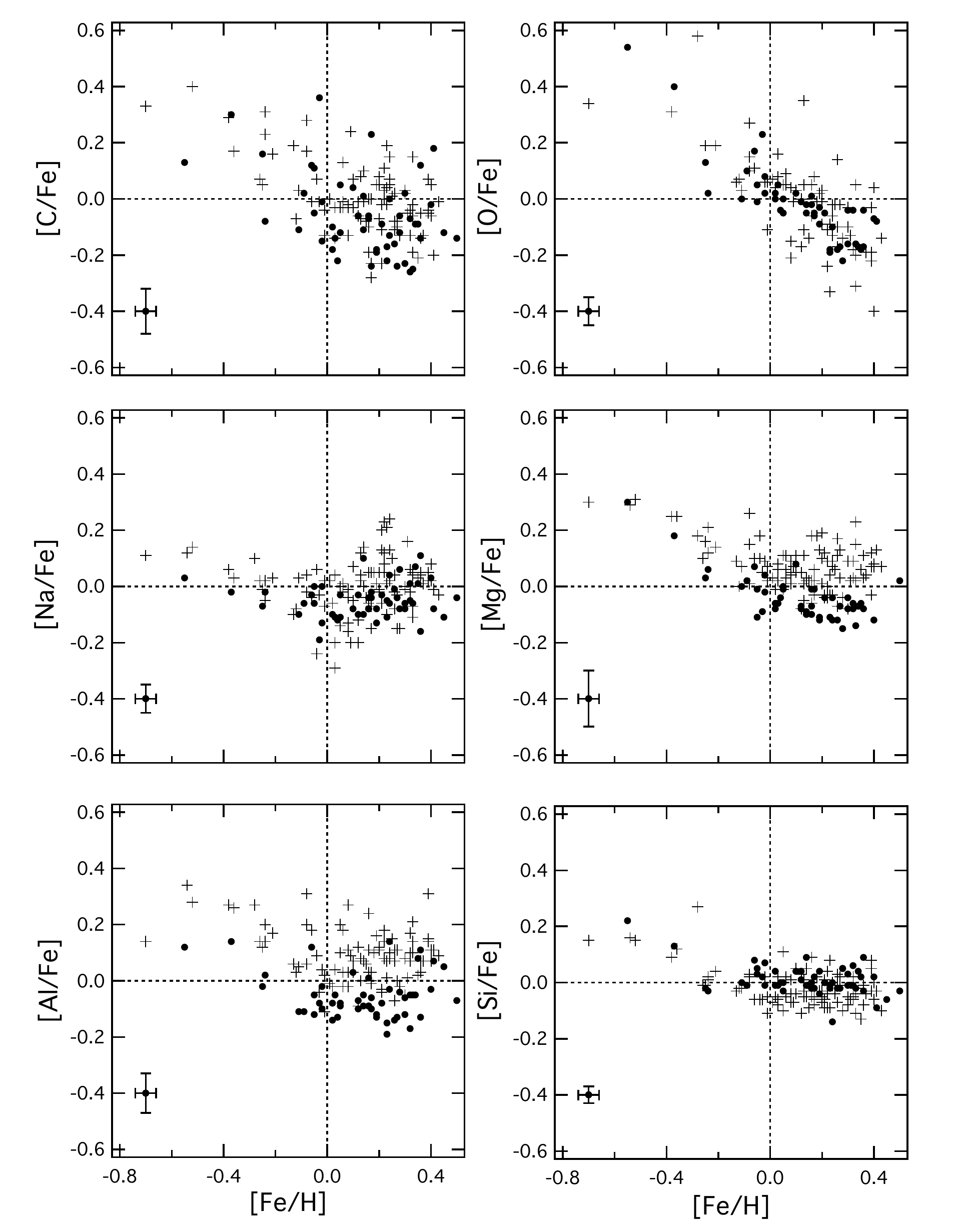}
 \caption{Values of [C/Fe], [O/Fe], [Na/Fe], [Mg/Fe], [Al/Fe] and [Si/Fe] versus [Fe/H] for SWPs from the present study (filled circles) and SWPs from \citet{gil06} (pluses). Typical error bars for the SWP abundance data from the present study are also shown.}
\end{figure}

\begin{figure}
  \includegraphics[width=3.5in]{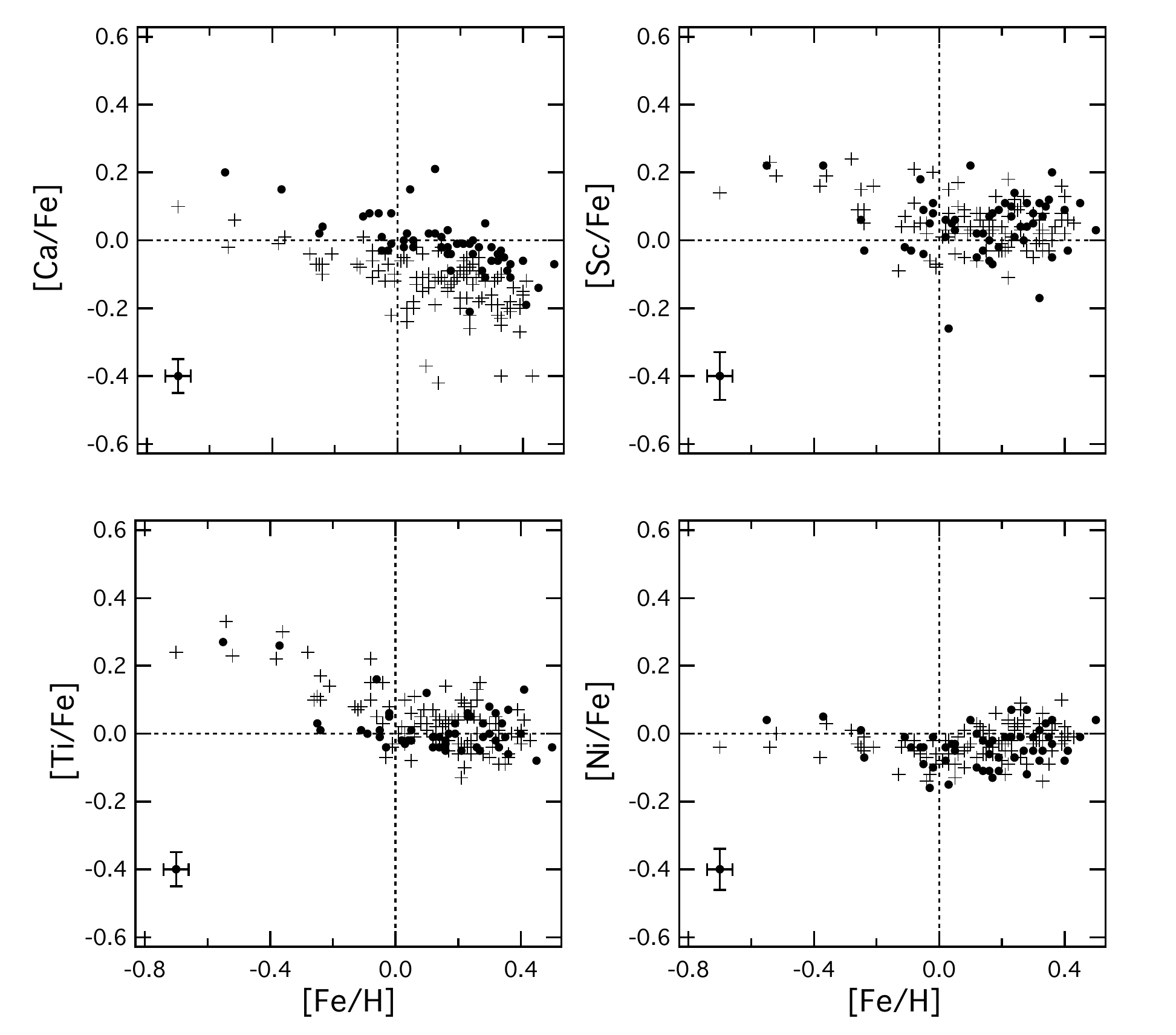}
 \caption{Same as Figure 2 but for [Ca/Fe], [Sc/Fe], [Ti/Fe] and [Ni/Fe].}
\end{figure}

\subsection{A Combined Sample}

The need for an extensive stellar control sample to compare to the SWPs has reinvigorated several research groups to more systematically characterize the physical parameters and chemical abundances of nearby sun-like stars. This environment, in which independent groups present estimates of basic physical data for the same stars, is an ideal one for ferreting out systematic differences among their data. 

Several research groups have reported evidence of anomalous abundance patterns among SWPs compared to stars without known planets. \citet{gg01} reported finding slightly smaller Na/Fe, Mg/Fe and Al/Fe abundance ratios among SWPs. \citet{bei05} measured these elements in 98 SWPs and 41 comparison stars and failed to find significant differences. However, the same group, upon expanding the samples to 101 SWPs and and 93 comparison stars, found smaller Al/Fe ratios and larger Mg/Fe ratios among the SWPs relative to the comparison stars \citep{gil06}. \citet{luck06} did not find any differences in the abundance patterns between their 55 SWPs relative to 161 comparison stars.

Employing a bootstrapping statistical analysis method, \citet{rob06} reported finding that the 99 SWPs in their Spectroscopic Properties of Cool Stars (SPOCS) database have significantly larger Si/Fe and Ni/Fe ratios than the 941 stars in their comparison sample. None of the other research groups have found anomalous values for these two elements among SWPs, but it is important to note that the analysis of \citet{rob06} is more sensitive than others.

In order to test these claims, we have constructed a new database of SWP and comparison star abundances by combining abundance data from several recent studies of these stars in a consistent way. The results of such a procedure are shown in Figures 4 to 7. We produced them in the following way.

\begin{figure}
  \includegraphics[width=3.5in]{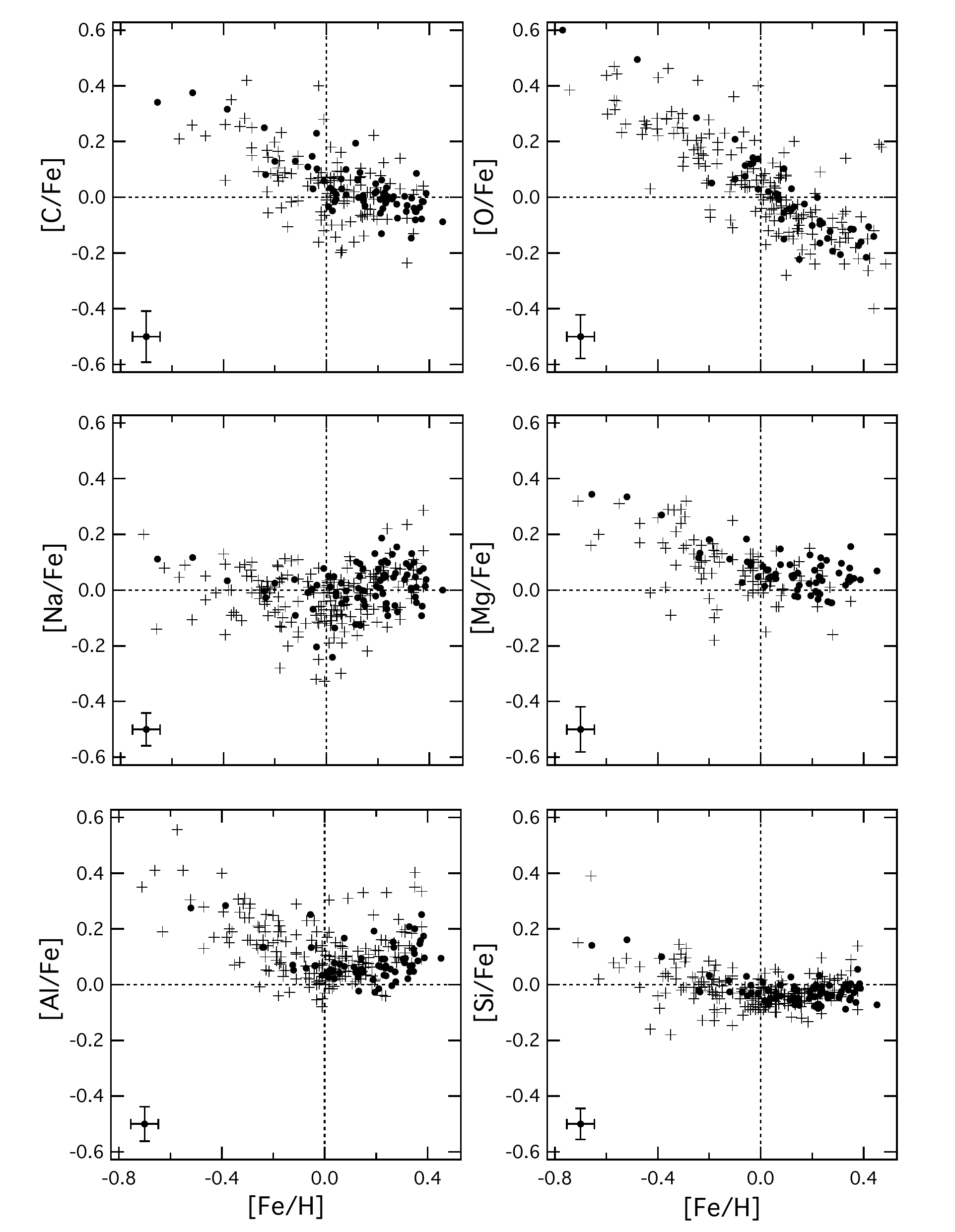}
 \caption{Values of [C/Fe], [O/Fe], [Na/Fe], [Mg/Fe], [Al/Fe] and [Si/Fe] versus [Fe/H] for SWPs corrected according to Equation 1 in the text using the constants in Tables 6 and 7. Comparison stars, corrected in the same way as the SWPs, are shown as pluses. The typical error bars for each element is shown on the lower left corner of each panel. See text for additional details.}
\end{figure}

\begin{figure}
  \includegraphics[width=3.5in]{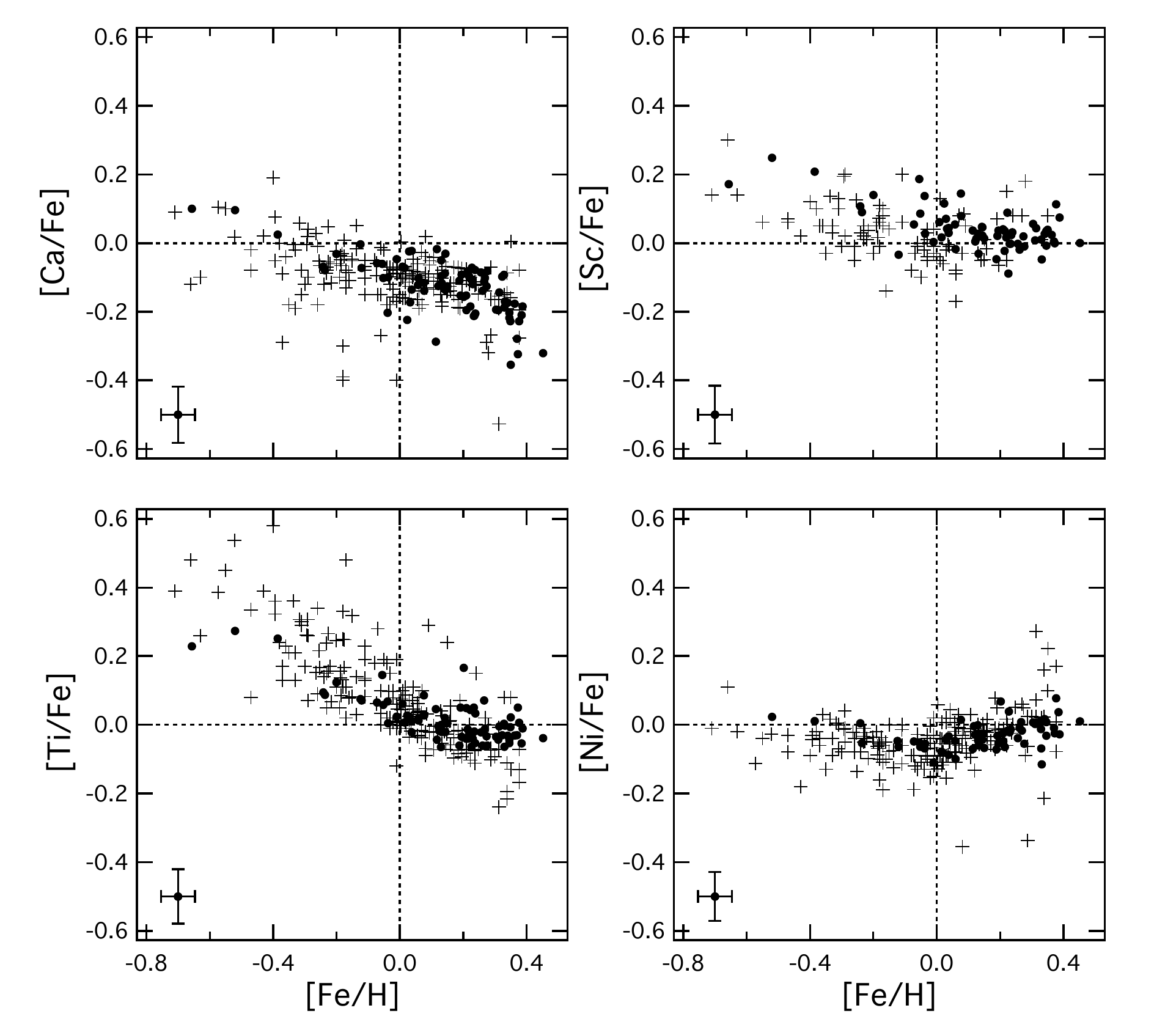}
 \caption{Same as Figure 4 but for [Ca/Fe], [Sc/Fe], [Ti/Fe] and [Ni/Fe].}
\end{figure}

We compiled SWP abundance data from \citet{bond06}, \citet{gil06}, \citet{gg00}, \citet{gg01}, the present work and \citet{luck06} for C, Na, Mg, Al, Si, Ca, Sc, Ti, Fe and Ni. For C, we also use the abundance results of \citet{ecu04} and \citet{takhon05}. The O abundances include data from Table 4 of \citet{ecu06b}, the present work, \citet{luck06} and \citet{takhon05}. We selected these particular studies due to their large sample sizes, the quality of their spectra and the similarities of their methods and temperature scales to each other. \citet{luck06} provide helpful intercomparisons of several recent large spectroscopic studies of nearby stars, including several of those employed here. We did not include in our analysis the data from the large Spectroscopic Properties of Cool Stars (SPOCS) database \citep{vf05} in order to test the claims of \citet{rob06} regarding anomalously high [Si/Fe] and [Ni/Fe] values in SWPs.

Next, we used data from the SWPs stars in common between pairs of studies listed above to calibrate out any systematic differences; data from comparison stars in common between \citet{gil06} and \citet{bond06} were also included. For a given pair of studies, we fit the abundance differences from the stars in common to the following equation:

\begin{equation}
$[el/H]$_{\rm 2} - $[el/H]$_{\rm 1} = A_{\rm 0} + A_{\rm 1} $T$_{\rm eff} + A_{\rm 2} $[el/H]$_{\rm 2}
\end{equation}
This equation allows us to correct for differences in zero-point, T$_{\rm eff}$ scale and [el/H] scale among the various studies. 

The values of T$_{\rm eff}$ are from \citet{gil06}, supplemented with a few values from \citet{bond06}. The values of [el/H]$_{\rm 1}$ are from \citet{gil06}. We list the derived values of the A$_{\rm 0}$, A$_{\rm 1}$ and A$_{\rm 2}$ constants in Tables 6 and 7. To apply the correction, the number calculated from Equation 1 is subtracted from [el/H]$_{\rm 2}$. In selecting the parameters to include in Equation 1, we also checked on possible correlations with $\log g$ but did not find anything significant. Thus, while one or more weak systematic trends may still exist in these data, we have little to gain at this point from including more terms in Equation 1.

We applied the calibration equations to the SWP data from the corresponding studies and calculated simple average abundances from the individual calibrated values for each element in each star. The number of calibration stars employed for each element ranged from about 30 to 50. In this way, we set \citet{gil06} as the reference for all the elements except C and O. For C we selected \citet{ecu04} as the reference, while for O we selected \citet{luck06}. Then, we calculated [el/Fe] values for each SWP using the calibrated values of [el/H] and [Fe/H]. Only SWPs with at least two measurements were retained. The final list contains 78 SWPs for all elements but C and O; there are 66 SWPs with C abundances and 60 SWPs with O abundances. 

We list the final corrected and combined abundance values for the SWPs in Table 8. For each element we also give the standard deviation of the abundance and the number of measurements. We calculated the standard deviation from the uncertainties from the individual studies summed in quadrature and averaged. This procedure probably slightly overestimates the uncertainties, but a calculation based on the scatter of the individual measurements for each star would underestimate them. The typical error bars we show in Figures 4 and 5 are based on the uncertainties listed in Table 8.

The abundance data for the comparison stars are from \citet{gil06} and \citet{bond06} for all elements but O.\footnote{The value of [Ca/H] of one of the comparison stars of \citet{bond06}, HD 193193, is listed by them as  -0.70. Jade Bond confirmed to us that this is a typo. The correct value is -0.07.} As in our analysis of the SWP abundances, we calibrated the comparison star abundances with Equation 1 and the constants listed in Tables 6 and 7. \citet{bond06} did not include Mg and Sc in their study, so we rely on \citet{gil06} exclusively for these element abundances for the comparison stars. Unlike the SWP data, however, we retained all the comparison stars after calibrating their abundances, since few of them are present in more than one study. Our final list contains 195 comparison stars for most elements. For O, the comparison stars are from \citet{luck06} and \citet{takhon05}, with \citet{luck06} set as the reference. The comparison stars for O number 180. 

While application of the abundance corrections introduces additional uncertainties (given the uncertainties in the coefficients to the equation), these are outweighed by the relative systematic abundance differences they remove from the various datasets. We did not add the additional uncertainties introduced from Equation 1 to the uncertainties due to the spectroscopic analyses, since our primary goal in the present study is to compare the abundance patterns of the SWPs and the comparison stars.

In Figures 6 and 7 we show binned [el/Fe] values resulting from application of Equation 1 and combining the data as described above. We binned the data as follows. For the SWPs, stars with [Fe/H] $< -0.20$ were included in the most metal-poor bin. The next bin included stars with $0.00 <$ [Fe/H] $\le -0.20$. The remaining bins are in steps of 0.1 dex in [Fe/H], but the most metal-rich one includes one star 0.05 dex beyond its range. The metal-rich bins contain about 10-20 stars each. For the comparison stars, the most-metal-poor bin includes stars with [Fe/H] $< -0.40$. The remaining bins are in steps of 0.1 dex in [Fe/H]. The number of stars i each bin ranges from about 15 to 40. 

The standard error of the mean, calculated from the scatter of data values within each bin, is also shown for each plotted point in Figures 6 and 7. This statistic may slightly underestimate the errors for some of the metal-poor bins, which contain relatively few stars.

\begin{figure}
  \includegraphics[width=3.5in]{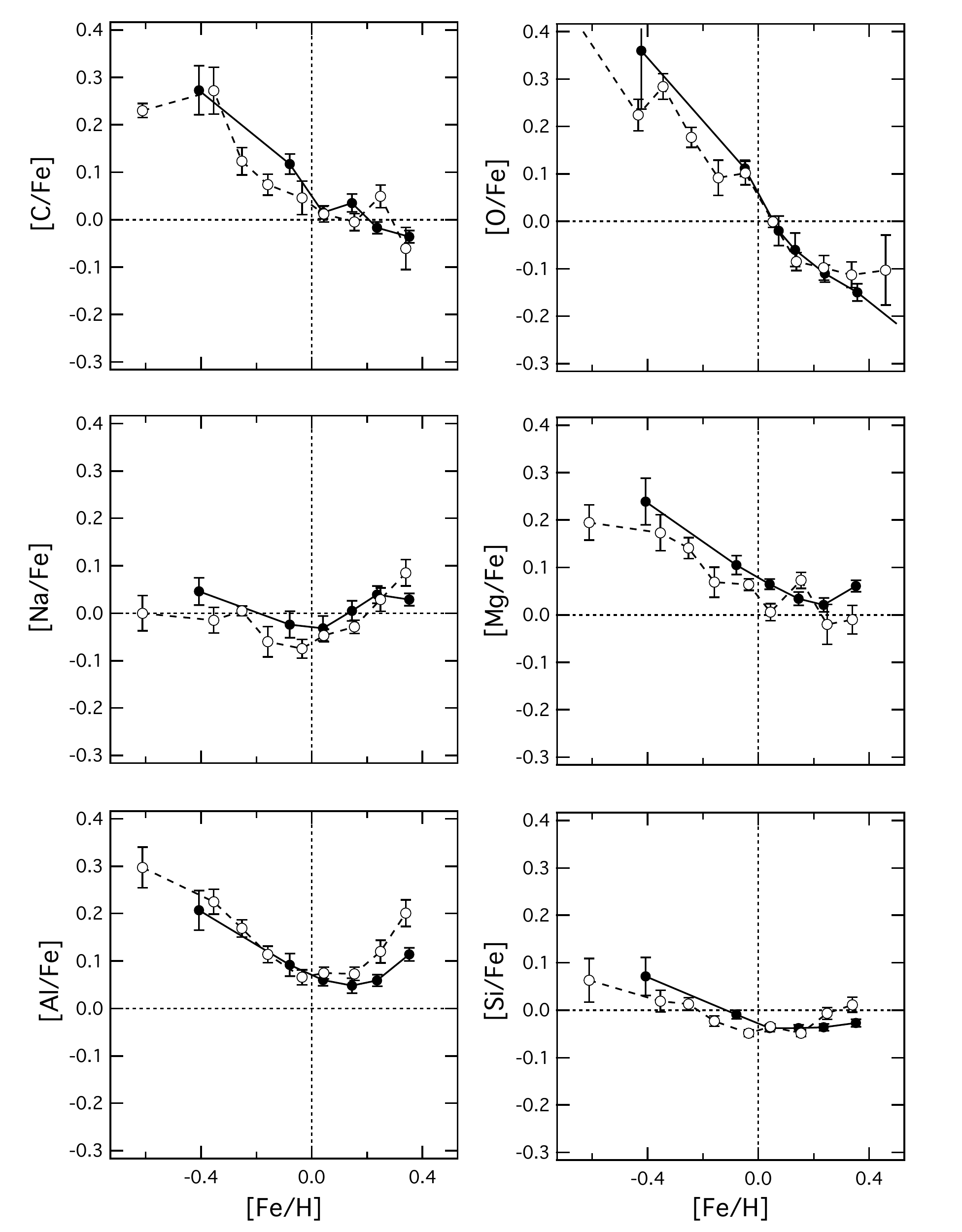}
 \caption{Binned values of [C/Fe], [O/Fe], [Na/Fe], [Mg/Fe], [Al/Fe] and [Si/Fe] versus [Fe/H] for SWPs (filled circles) and control stars (open circles).}
\end{figure}

\begin{figure}
  \includegraphics[width=3.5in]{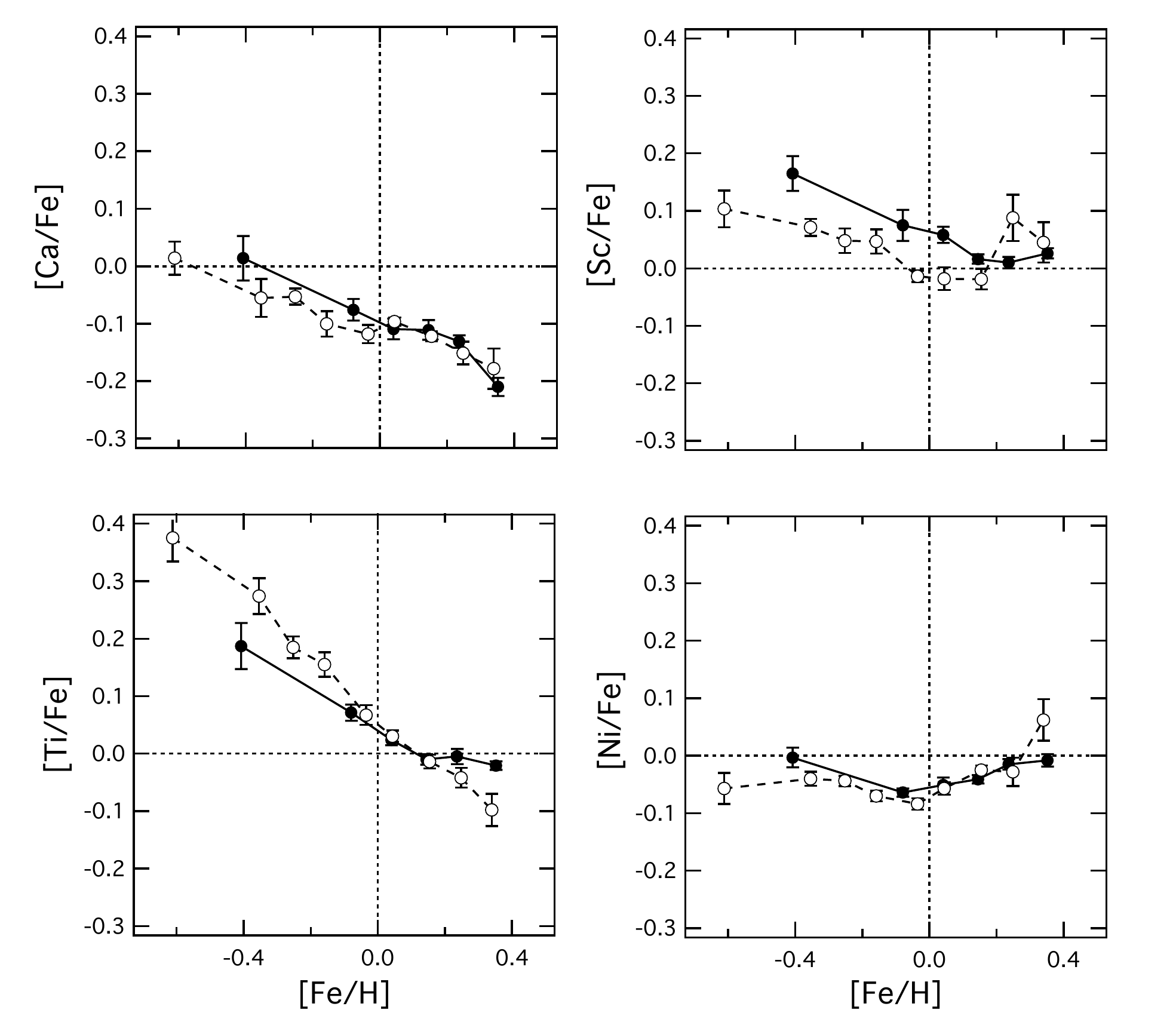}
 \caption{Binned values of [Ca/Fe], [Sc/Fe], [Ti/Fe] and [Ni/Fe] versus [Fe/H] for SWPs (filled circles) and comparison stars (open circles).}
\end{figure}

\section{Discussion}

Comparison of Figures 2 and 3 to Figures 4 and 5 confirms that our data correction and combination procedure reduced the scatter in the data points. For example, the standard deviations of our [Mg/Fe], [Al/Fe] and [Si/Fe] values in Figure 2 in the [Fe/H] range from 0.00 to 0.20 are $\pm$ 0.051, 0.045 and 0.033 dex, respectively; the corresponding values for the \citet{gil06} SWP data are $\pm$ 0.061, 0.073 and 0.053 dex. The corresponding numbers for the corrected and combined data in Figure 4 are $\pm$ 0.044, 0.049 and 0.026 dex, respectively.

This improved dataset allows us to search for subtle differences between the SWPs and the comparison stars. The metal-poor bins are well-represented by the comparison stars, but there are very few SWPs with [Fe/H] $< -0.2$. Furthermore, the binning is different for [Fe/H] $< 0.00$. Therefore, we restrict our comparisons to data with [Fe/H] greater than about $-0.1$.

Visual inspection of the binned data in Figures 6 and 7 reveals several differences between the SWP and comparison star abundances. Most of the elements plotted in the figures display at least one significant difference in the binned data. More significant are those cases that display different trends among the metal-rich bins. Among the most metal-rich stars, for example, the Al/Fe and Si/Fe abundance ratios are systematically smaller for the SWPs than for the comparison stars, and the Ti/Fe ratio displays the opposite trend. Other abundance ratios, such as Na/Fe, Mg/Fe, Sc/Fe and Ni/Fe, display more subtle differences in the trends. The differences appear smallest for the C/Fe, O/Fe and Ca/Fe abundance ratios.

We show in Figure 8 the binned abundances for the elements Na, Si, Ti and Ni from the SPOCS database. The data were binned as in Figures 6 and 7. The SWPs number 104 and the comparison stars number 788; we excluded giant stars from the samples. Visual inspection of the trends in Figure 8 shows that they are generally similar to those in Figures 6 and 7 for the same elements. For all four elements shown in Figure 8 (but most prominently for Si and Ni), the SWPs have higher abundance ratios than the comparison stars at about [Fe/H] $= +0.25$ dex. This is consistent with the findings of \citet{rob06}. At the highest value of [Fe/H], [Na/Fe], [Si/Fe] and [Ti/Fe] are smaller for the SWPs, but the error bars are larger at this [Fe/H] bin.

\begin{figure}
  \includegraphics[width=3.5in]{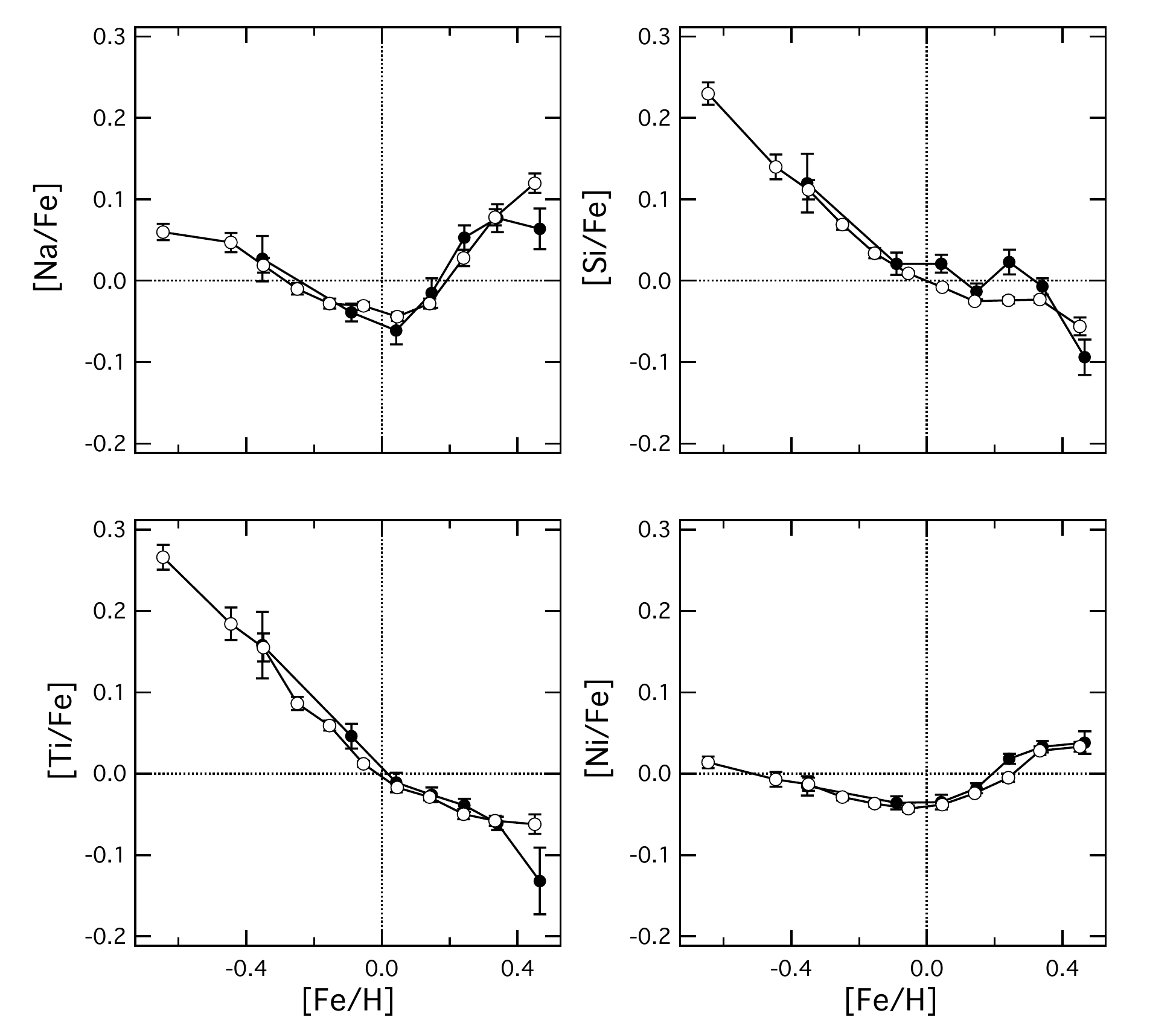}
 \caption{Binned values of [Na/Fe], [Si/Fe], [Ti/Fe] and [Ni/Fe] versus [Fe/H] for SWPs (filled circles) and comparison stars (open circles) from the SPOCS database.}
\end{figure}

The data on Na, Si, Ti and Ni abundances from the SPOCS database are not consistent with our findings. The statistical analysis of \citet{rob06} indicates that SWPs have significantly more Si and Ni than stars without planets and that Ti is not significantly different between the two samples.

It is not clear which of these results is correct. \citet{vf05} employed a new method of abundance analysis very different from the more traditional method employed by the other recent spectroscopic studies of SWPs. On the face of it, we should expect the random errors to be smaller for the comparison stars sample in the SPOCS database, given its size and homogeneity, but the number of SWPs in SPOCS is similar to that in \citet{gil06}.

The implications of real differences in the chemical abundance patterns of SWPs compared to stars without planets are far-ranging. For example, \citet{gg06} employed some of the abundance data from the present work, along with data from other studies, to search for anomalous abundance trends with condensation temperature. No significant trends were found; \citet{ecu06a} and \citet{huang} independently reached the same conclusions. Had such trends been found, they would have indicated that accretion of metal-rich material onto the convection zones of SWPs ('self-enrichment') may account for at least part of their high observed metallicities \citep{ggpasp06}. Nevertheless, the search for trends with condensation temperature should continue as the sample of SWPs continues to grow; even upper limits on the amount of accreted metal-rich material can be helpful.

The chemical abundance patterns among SWPs can also give us clues about giant planet formation processes. For example, the absence of strong evidence for competing explanations for the planet-metallicity correlation leads us to conclude that the 'primordial' explanation accounts for the bulk of the data. In addition, \citet{idalin05} have succeeded in reproducing the observed planet-metallicity correlation within the theoretical framework of the core instability accretion model. 

\citet{rob06} examined the abundance patterns in SWPs to determine if initial metallicity is the only controlling compositional factor in giant planet formation. If giant planet formation depends on a nearby supernova, for example, then not only will the metallicities differ but so will the chemical abundance patterns of SWPs differ from those of other nearby stars. It could be the case that an element other than Fe (employed as the primary metallicity indicator) is physically more important for planet formation. If overall metallicity is the only important factor, then the el/Fe ratios should be the same in SWPs and comparison stars. 

If an element other than Fe is important for planet formation, then differences between SWPs and comparison stars will be evident in plots such as those shown in Figures 6 and 7. Among metal-rich stars there is a spread in [el/Fe] of a few tenths of a dex at a given value of [Fe/H]. Some of it is due to measurement error, but some is intrinsic. The intrinsic spread in the abundance of an element other than Fe critical to giant planet formation could be sufficient to account for the observed planet-metallicity correlation. In this case the Fe abundance would not be the primary determinant of giant planet formation; \citet{rob06} suggest in this case that ``...iron abundance derives its power as a predictor of planet presence from its correlation with the abundance of another element of more physical importance to the planet formation process ....''

Based on their finding higher [Si/Fe] values among SWPs, \citet{rob06} predicted that other $\alpha$-elements would also be found to be enhanced in SWPs. In particular, they predicted C, O, Ne, Mg, S and Ar should also be enhanced in SWPs; since O is the most abundant element at the location in the protoplanetary disk where giant planets form, they predict it should be more enhanced in SWPs than the other elements. We do not confirm their prediction of higher [O/Fe] values in SWPs.

\citet{ggpasp06} gave several suggestions for improving abundance analyses of SWPs and comparison stars. In particular, there is need for more comparison stars with [Fe/H] $\ge 0.3$ dex. The SWPs data extend to nearly 0.4 dex. Also, the number of SWPs with [Fe/H] values below -0.2 dex is relative small. This situation will improve as the number of known SWPs continues to grow. Finally, it should be relatively straightforward to double the number of SWPs with multiple independent spectroscopic abundance analyses, given the number of groups involved in such work. Of particular importance will be improvements in O abundance data.

\section{Conclusions}

We have determined [el/H] values for 18 elements in the SWPs studied by \citet{laws03}: Li, C, N, O, Na, Mg, Al, Si, S, Ca, Sc, Ti, Cr, Mn, Ni, Cu, Zn and Eu. Comparison of our results to those of other recent spectroscopic studies of SWPs reveals small abundance offsets among them.

We combined our abundance data with similar data from other studies of SWPs for a subset of these elements. Prior to combining the datasets, we corrected the individual abundance values for systematic differences among the studies. The resulting combined abundances exhibit  slightly different Al/Fe, Si/Fe and Ti/Fe ratios between the metal-rich SWPs and comparison stars; more subtle differences are apparent for Na/Fe, Mg/Fe, Sc/Fe and Ni/Fe. We do not confirm the findings of \citet{rob06} that Si and Ni abundances are higher in SWPs compared to stars without planets. Neither do we confirm their prediction that O abundances should be higher among SWPs. We encourage additional studies of the abundance patterns of SWPs with the goal of resolving these differing results.

\begin{table*}
\centering
\caption{[X/H] values for Li, C, N and O.}

\label{xmm}
\begin{tabular}{lccccccc}
\hline
Star & Li & C & C & C & N & O & O\\
(HD) & & (EWs) & (synth) & (avg) & & (trip) & (corr)\\
\hline
4203 & $-0.20 \pm 0.08$ & $~~0.45 \pm 0.06$ & $~~0.30 \pm 0.07$ & $~~0.38 \pm 0.08$ & $~~0.27 \pm 0.10$ & $~~0 .35 \pm 0.04$ & $~~0.33 \pm 0.04$\\
4208 & $-0.55 \pm 0.10$ & $-0.30 \pm 0.04$ & $~~0.11 \pm 0.10$ & $-0.09 \pm 0.09$ & --- & $-0.17 \pm 0.02$ & $-0.12 \pm 0.02$\\
6434 & $-0.30 \pm 0.10$ & $-0.40 \pm 0.05$ & $-0.43 \pm 0.10$ & $-0.42 \pm 0.10$ & --- & $-0.01 \pm 0.03$ & $-0.01 \pm 0.03$\\
8574 & $~~1.35 \pm 0.04$ & $-0.10 \pm 0.08$ & $-0.18 \pm 0.05$ & $-0.17 \pm 0.08$ & $-0.37 \pm 0.10$ & $~~0.02 \pm 0.13$ & $~~0.00 \pm 0.13$\\
16141 & $~~0.19 \pm 0.15$ & $~~0.00 \pm 0.03$ & $~~0.02 \pm 0.08$ & $~~0.01 \pm 0.07$ & $-0.09 \pm 0.10$ & $~~0.13 \pm 0.05$ & $~~0.10 \pm 0.05$\\
19994 & $~~0.84 \pm 0.05$ & $~~0.08 \pm 0.06$ & $-0.03 \pm 0.08$ & $~~0.03 \pm 0.09$ & $-0.06 \pm 0.10$ & $~~0.20 \pm 0.04$ & $~~0.09 \pm 0.04$\\
22049 & $-0.79 \pm 0.10$ & $-0.10 \pm 0.21$ & $~~0.02 \pm 0.10$ & $-0.07 \pm 0.20$ & --- & $-0.09 \pm 0.01$ & $~~0.01 \pm 0.01$\\
27442 & $-0.99 \pm 0.10$ & $~~0.82 \pm 0.73$ & $~~0.36 \pm 0.10$ & $~~0.59 \pm 0.65$ & $~~0.78 \pm 0.10$ & $~~0.25 \pm 0.03$ & $~~0.33 \pm 0.03$\\
28185 & $ -0.20 \pm 0.10$  & $~~0.18 \pm 0.02$ & $~~0.03 \pm 0.10$ & $~~0.11 \pm 0.09$ & $~~0.06 \pm 0.10$  & $~~0.13 \pm 0.03$  & $~~0.14 \pm 0.03$ \\
33636 & $~~1.36 \pm 0.04$  & $ -0.30 \pm 0.04$ & $ -0.09 \pm 0.05$ & $-0.22 \pm 0.06$ & $ -0.15 \pm   0.10$  & $-0.11 \pm 0.03$  & $ -0.11 \pm 0.03$ \\
37124 & $ -0.47 \pm 0.15$  & $ -0.20 \pm 0.03$ & $~~0.13 \pm 0.10$ & $-0.07 \pm 0.09$ & $~~0.29 \pm   0.10$  & $~~0.00 \pm 0.05$  & $~~0.03 \pm 0.05$ \\
46375 & $ -0.54 \pm 0.10$  & $~~0.31 \pm 0.14$ & $~~0.32 \pm 0.10$ & $~~0.32 \pm 0.15$ & $~~0.41 \pm 0.10$  & $~~0.20 \pm 0.04$  & $~~0.26 \pm 0.04$ \\
50554 & $~~1.35 \pm 0.10$  & $ -0.10 \pm 0.05$ & $ -0.01 \pm 0.10$ & $-0.08 \pm 0.10$ & $~~0.02 \pm   0.10$  & $~~0.08 \pm 0.04$  & $~~0.04 \pm 0.04$ \\
68988 & $~~0.99 \pm 0.05$  & $~~0.26 \pm 0.02$ & $~~0.24 \pm 0.05$ & $~~0.25 \pm 0.05$ & $~~0.24 \pm 0.10$  & $~~0.23 \pm 0.04$  & $~~0.17 \pm 0.04$ \\
82943 & $~~1.31 \pm 0.03$  & $~~0.03 \pm 0.04$ & $~~0.16 \pm 0.05$ & $~~0.10 \pm 0.06$ & $~~0.02 \pm 0.10$  & $~~0.13 \pm 0.03$  & $~~0.08 \pm 0.03$ \\
83443 & $~~0.01 \pm 0.12$  & $~~0.50 \pm 0.14$ & $~~0.46 \pm 0.05$ & $~~0.48 \pm 0.13$ & $~~0.51 \pm 0.10$  & $~~0.30 \pm 0.03$  & $~~0.32 \pm 0.03$ \\
95128 & $~~0.55 \pm 0.05$  & $~~0.00 \pm 0.06$ & $~~0.20 \pm 0.05$ & $~~0.10 \pm 0.07$ & $ -0.15 \pm 0.10$  & $~~0.08 \pm 0.04$  & $~~0.05 \pm 0.04$ \\
106252 & $~~0.65 \pm 0.05$  & $ -0.10 \pm 0.03$ & $~~0.26 \pm 0.10$ & $~~0.06 \pm 0.09$ & $ -0.24 \pm 0.10$  & $~~0.01 \pm 0.04$  & $~~0.00 \pm 0.04$ \\
108147 & $~~1.23 \pm 0.05$  & $~~0.00 \pm 0.07$ & $~~0.15 \pm 0.07$ & $~~0.06 \pm 0.09$ & $   -0.42 \pm 0.10$  & $~~0.15 \pm 0.03$  & $~~0.05 \pm 0.03$ \\
114783 & $ -1.06 \pm 0.08$  & $~~0.22 \pm 0.11$ & $~~0.57 \pm 0.12$ & $~~0.40 \pm 0.14$ & $~~0.23 \pm 0.10$  & $~~0.03 \pm 0.03$  & $~~0.12 \pm 0.03$ \\
117176 & $~~0.68 \pm 0.05$  & $~~0.00 \pm 0.04$ & $~~0.03 \pm 0.10$ & $-0.03 \pm 0.09$ & $ -0.24 \pm 0.10$  & $~~0.04 \pm 0.03$  & $~~0.06 \pm 0.03$ \\
121504 & $~~1.37 \pm 0.04$  & $~~0.00 \pm 0.09$ & $~~0.17 \pm 0.07$ & $~~0.06 \pm 0.10$ & $   -0.08 \pm 0.10$  & $~~0.16 \pm 0.08$  & $~~0.11 \pm 0.08$ \\
136118 & $~~1.20 \pm 0.04$  & $ -0.10 \pm 0.10$ & $ -0.10 \pm 0.10$ & $-0.10 \pm 0.12$ & $ -0.68 \pm   0.10$  & $~~0.03 \pm 0.05$  & $ -0.06 \pm 0.05$ \\
141937 & $~~1.24 \pm 0.05$  & $~~0.00 \pm 0.08$ & $~~0.34 \pm 0.05$ & $~~0.15 \pm 0.08$ & $   -0.21 \pm 0.10$  & $~~0.15 \pm 0.06$  & $~~0.12 \pm 0.06$ \\
160691 & $ -0.17 \pm 0.10$  & $~~0.16 \pm 0.05$ & $~~0.28 \pm 0.10$ & $~~0.22 \pm 0.10$ & $~~0.37 \pm 0.10$  & $~~0.07 \pm 0.03$  & $~~0.06 \pm 0.03$ \\
168746 & $ -0.36 \pm 0.15$  & $~~0.00 \pm 0.06$ & $~~0.19 \pm 0.08$ & $~~0.06 \pm 0.09$ & $ -0.16 \pm 0.10$  & $~~0.09 \pm 0.02$  & $~~0.11 \pm 0.02$ \\
169830 & $~~0.07 \pm 0.10$  & $ -0.10 \pm 0.09$ & $~~0.01 \pm 0.07$ & $-0.07 \pm 0.10$ & $~~0.25 \pm 0.10$  & $~~0.24 \pm 0.02$  & $~~0.11 \pm 0.02$ \\
190228 & $~~0.06 \pm 0.10$  & $ -0.40 \pm 0.09$ & $ -0.22 \pm 0.07$ & $-0.32 \pm 0.10$ & $ -0.39 \pm   0.10$  & $-0.30 \pm 0.04$  & $ -0.22 \pm 0.04$ \\
195019 & $~~0.23 \pm 0.10$  & $~~0.00 \pm 0.03$ & $ -0.14 \pm 0.05$ & $-0.11 \pm 0.05$ & $ -0.33 \pm 0.10$  & $~~0.09 \pm 0.04$  & $~~0.08 \pm 0.04$ \\
202206 & $~~0.48 \pm 0.05$  & $~~0.12 \pm 0.03$ & $~~0.04 \pm 0.08$ & $~~0.08 \pm 0.07$ & $~~0.19 \pm 0.10$  & $~~0.18 \pm 0.03$  & $~~0.17 \pm 0.03$ \\
213240 & $~~1.36 \pm 0.03$  & $~~0.00 \pm 0.03$ & $~~0.03 \pm 0.12$ & $~~0.01 \pm 0.11$ & $~~0.00 \pm 0.10$  & $~~0.10 \pm 0.06$  & $~~0.04 \pm 0.06$ \\
\hline
\end{tabular}
\end{table*}

\begin{table*}
\centering
\caption{[X/H] values for Al, Ca, Mg, Na and S.}

\label{xmm}
\begin{tabular}{lccccccc}
\hline
Star & Al & Al & Al & Ca & Mg & Na & S \\
(HD) & (EWs) & (synth) & (avg) & & & & \\
\hline
4203 & $~~0.47 \pm 0.05$  & $~~0.27 \pm 0.05$ & $~~0.37 \pm 0.06$ & $~~0.34 \pm 0.05$ & $~~0.28 \pm 0.10$  & $~~0.43 \pm 0.07$  & $~~0.42 \pm 0.06$ \\
4208 & $ -0.32 \pm 0.03$  & $ -0.20 \pm 0.05$ & $ -0.27 \pm 0.05$ & $-0.23 \pm 0.03$ & $ -0.22 \pm  0.10$  & $-0.32 \pm 0.02$  & $ -0.15 \pm 0.13$ \\
6434 & $ -0.46 \pm 0.02$  & $ -0.40 \pm 0.05$ & $ -0.43 \pm 0.05$ & $-0.35 \pm 0.03$ & $ -0.25 \pm  0.10$  & $-0.52 \pm 0.02$  & $-0.58 \pm 0.09$ \\
8574 & $ -0.06 \pm 0.04$  & $ -0.10 \pm 0.05$ & $ -0.12 \pm 0.06$ & $~~0.06 \pm 0.04$ & $~~0.02 \pm   0.10$  & $-0.02 \pm 0.06$  & $ -0.11 \pm 0.05$ \\
16141 & $~~0.07 \pm 0.02$  & $~~0.07 \pm 0.04$ & $~~0.07 \pm 0.04$ & $~~0.18 \pm 0.03$ & $~~0.08 \pm 0.10$ & $~~0.06 \pm 0.05$  & $~~0.06 \pm 0.05$ \\
19994 & $~~0.12 \pm0.02$  & $~~0.05 \pm 0.04$ & $~~0.09 \pm 0.04$ & $~~0.15 \pm 0.07$ & $~~0.04 \pm 0.10$  & $~~0.24 \pm 0.04$  & $~~0.05 \pm 0.10$ \\
22049 & $ -0.18 \pm 0.04$  & $ -0.20 \pm 0.07$ & $ -0.20 \pm 0.07$ & $-0.01 \pm 0.06$ & $ -0.07 \pm 0.10$  & $-0.15 \pm 0.04$  & $~~0.03 \pm 0.08$ \\
27442 & $~~0.51 \pm 0.06$ & $~~0.46 \pm 0.06$ & $~~0.48 \pm 0.07$ & $~~0.22 \pm 0.06$ & ---  & $~~0.33 \pm 0.02$ & $~~0.86 \pm 0.11$ \\
28185 & $~~0.21 \pm0.06$  & $~~0.21 \pm 0.06$ & $~~0.21 \pm 0.07$ & $~~0.20 \pm 0.07$ & $~~0.12 \pm 0.10$  & $~~0.28 \pm 0.07$  & $~~0.16 \pm 0.07$ \\
33636 & $ -0.23 \pm 0.02$  & $ -0.20 \pm 0.08$ & $ -0.22 \pm 0.07$ & $-0.04 \pm 0.04$ & $ -0.11 \pm 0.10$  & $ -0.21 \pm 0.02$  & $ -0.31 \pm 0.05$ \\
37124 & $ -0.35 \pm 0.02$  & $ -0.10 \pm 0.06$ & $ -0.23 \pm 0.05$ & $-0.22 \pm 0.08$ & $ -0.19 \pm 0.10$  & $ -0.39 \pm 0.04$  & $ -0.22 \pm 0.12$ \\
46375 & $~~0.21 \pm 0.04$  & $~~0.28 \pm 0.06$ & $~~0.24 \pm 0.06$ & $~~0.24 \pm 0.04$ & $~~0.22 \pm 0.10$  & $~~0.22 \pm 0.08$  & $~~0.47 \pm 0.04$ \\
50554 &$ -0.12 \pm 0.04$  & $ -0.10 \pm 0.04$ & $ -0.12 \pm 0.05$ & $~~0.02 \pm 0.04$ & $ -0.06 \pm 0.10$  & $ -0.08 \pm 0.04$  & $ -0.19 \pm 0.07$ \\
68988 & $~~0.33 \pm 0.02$  & $~~0.24 \pm 0.05$ & $~~0.29 \pm 0.05$ & $~~0.29 \pm 0.03$ & $~~0.27 \pm 0.10$ & $~~0.41 \pm 0.05$  & $~~0.41 \pm 0.13$ \\
82943 & $~~0.10 \pm 0.02$  & $~~0.13 \pm 0.05$ & $~~0.12 \pm 0.04$ & $~~0.24 \pm 0.04$ & $~~0.14 \pm 0.10$  & $~~0.25 \pm 0.03$  & $~~0.10 \pm 0.02$ \\
83443 & $~~0.52 \pm 0.06$  & $~~0.41 \pm 0.06$ & $~~0.47 \pm 0.07$ & $~~0.29 \pm 0.05$ & ---  & $~~0.47 \pm 0.03$ & $~~0.58 \pm 0.08$ \\
95128 & $ -0.01 \pm 0.06$  & $~~0.00 \pm 0.06$ & $ -0.03 \pm 0.07$ & $ ~~0.03 \pm 0.04$ & $~~0.05 \pm 0.10$  & $~~0.02 \pm 0.04$  & $ -0.05 \pm 0.10$ \\
106252 & $ -0.14 \pm 0.03$  & $~~0.00 \pm 0.05$ & $ -0.10 \pm 0.05$ & $-0.08 \pm 0.02$ & $ -0.06 \pm 0.10$  & $ -0.11 \pm 0.02$  & $ -0.07 \pm 0.02$ \\
108147 & $~~0.13 \pm 0.04$  & $~~0.03 \pm 0.05$ & $~~0.08 \pm 0.06$ & $~~0.22 \pm 0.04$ & $~~0.12 \pm 0.10$  & $~~0.12 \pm 0.04$  & $ -0.06 \pm 0.03$ \\
114783 & $~~0.09 \pm 0.06$  & $~~0.04 \pm 0.07$ & $~~0.07 \pm 0.08$ & $~~0.13 \pm 0.08$ & $~~0.16 \pm 0.10$  & $~~0.15 \pm 0.06$  & $~~0.21 \pm 0.12$ \\
117176 & $ -0.06 \pm 0.03$  & $~~0.00 \pm 0.08$ & $ -0.04 \pm 0.07$ & $ -0.03 \pm 0.04$ & $ -0.04 \pm 0.10$  & $ -0.15 \pm 0.04$  & $ -0.11 \pm 0.10$ \\
121504 & $~~0.11 \pm 0.02$  & $~~0.00 \pm 0.06$ & $~~0.05 \pm 0.05$ & $~~0.33 \pm 0.19$ & $~~0.05 \pm 0.10$  & $~~0.02 \pm 0.02$  & $~~0.16 \pm 0.13$ \\
136118 & $ -0.13 \pm 0.02$  & $ -0.20 \pm 0.07$ & $ -0.17 \pm 0.06$ & $ -0.04 \pm 0.08$ & $ -0.16 \pm 0.10$  & $-0.05 \pm 0.03$  & $ -0.18 \pm 0.05$ \\
141937 & $~~0.00 \pm 0.03$  & $~~0.09 \pm 0.07$ & $~~0.05 \pm 0.07$ & $~~0.12 \pm 0.04$ & $~~0.05 \pm 0.10$  & $~~0.04 \pm 0.03$ & $ -0.08 \pm 0.06$ \\
160691 & $~~0.30 \pm 0.08$  & $~~0.18 \pm 0.06$ & $~~0.24 \pm 0.09$ & $~~0.17 \pm 0.02$ & ---  & $~~0.34 \pm 0.07$  & $~~0.26 \pm 0.03$ \\
168746 & $~~0.03 \pm 0.02$  & $~~0.09 \pm 0.05$ & $~~0.06 \pm 0.05$ & $~~0.02 \pm 0.04$ & $~~0.01 \pm 0.10$  & $ -0.09 \pm 0.02$  & $ -0.15 \pm 0.06$ \\
169830 & $~~0.14 \pm 0.03$  & $~~0.07 \pm 0.07$ & $~~0.11 \pm 0.07$ & $~~0.08 \pm 0.05$ & ---  & $~~0.13 \pm 0.06$  & $ -0.02 \pm 0.03$ \\
190228 & $ -0.25 \pm 0.02$  & $ -0.10 \pm 0.06$ & $ -0.22 \pm 0.06$ & $ -0.20 \pm 0.05$ & $ -0.18 \pm 0.10$ & $-0.26 \pm 0.03$  & $ -0.31 \pm 0.08$ \\
195019 & $ -0.03 \pm 0.02$  & $~~0.00 \pm 0.05$ & $ -0.02 \pm 0.05$ & $ ~~0.05 \pm 0.05$ & $ -0.03 \pm 0.10$  & $ -0.08 \pm 0.03$  & $ -0.08 \pm 0.05$ \\
202206 & $~~0.23 \pm 0.04$ & $~~0.33 \pm 0.05$ & $~~0.28 \pm 0.06$ & $~~0.30 \pm 0.03$ & $~~0.19 \pm 0.10$  & $~~0.27 \pm 0.04$  & $~~0.24 \pm 0.07$ \\
213240 & $ -0.03 \pm 0.02$ & $~~0.12 \pm 0.05$ & $~~0.04 \pm 0.04$ & $~~0.02 \pm 0.21$ & ---  & $~~0.18 \pm 0.04$ & $ -0.30 \pm 0.11$ \\
\hline
\end{tabular}
\end{table*}

\begin{table*}
\centering
\caption{[X/H] values for Sc, Si and Ti.}

\label{xmm}
\begin{tabular}{lccccc}
\hline
Star & Sc & Si & Ti I & Ti II & Ti \\
(HD) & & & & & (avg) \\
\hline
4203 & $~~0.49 \pm 0.02$  & $~~0.42 \pm 0.04$ & $~~0.43 \pm 0.07$  & $~~0.37 \pm  0.04$  & $~~0.40 \pm 0.07$ \\
4208 & $ -0.19 \pm 0.06$ & $ -0.27 \pm 0.02$ & $ -0.21 \pm 0.03$ & $ -0.23 \pm  0.07$  & $ -0.22 \pm 0.07$ \\
6434 & $ -0.33 \pm 0.08$  & $ -0.33 \pm 0.02$ & $ -0.32 \pm 0.02$  & $ -0.25 \pm 0.04$  &$ -0.28 \pm 0.04$ \\
8574 & $~~0.09 \pm 0.08$ & $~~0.05 \pm 0.02$ & $~~0.01 \pm 0.04$ & $~~0.08 \pm 0.10$ & $~~0.04 \pm 0.09$ \\
16141 & $~~0.28 \pm 0.05$  & $~~0.15 \pm 0.02$ & $~~0.17 \pm 0.02$ & $~~0.26 \pm  0.03$ & $~~0.22 \pm 0.03$ \\
19994 & $~~0.11 \pm 0.02$ & $~~0.23 \pm 0.02$ & $~~0.16 \pm 0.02$ & $~~0.04 \pm 0.02$ & $~~0.10 \pm 0.02$ \\
22049 & $ -0.12 \pm 0.07$ & $ -0.10 \pm 0.04$ & $~~0.03 \pm 0.07$ & $ -0.20 \pm 0.06$ & $ -0.09 \pm 0.08$ \\
27442 & $~~0.38 \pm 0.10$ & $~~0.32 \pm 0.12$ & $~~0.54 \pm 0.22$ & --- & $~~0.54 \pm 0.22$ \\
28185 & $~~0.38 \pm 0.04$ & $~~0.10 \pm 0.16$ & $~~0.28 \pm 0.03$ & $~~0.30 \pm 0.02$ & $~~0.29 \pm 0.03$ \\
33636 & $ -0.13 \pm 0.06$ & $ -0.11 \pm 0.11$ & $ -0.08 \pm 0.02$ & $ -0.13 \pm  0.10$ & $ -0.10 \pm 0.09$ \\
37124 & $ -0.15 \pm 0.08$ & $ -0.24 \pm 0.02$ & $ -0.12 \pm 0.04$ & $ -0.09 \pm  0.05$ & $ -0.11 \pm 0.06$ \\
46375 & $~~0.38 \pm 0.04$ & $~~0.33 \pm 0.05$ & $~~0.45 \pm 0.11$ & $~~0.31 \pm 0.02$ & $~~0.38 \pm 0.10$ \\
50554 & $~~0.03 \pm 0.08$ & $~~0.01 \pm 0.02$ & $ -0.02 \pm 0.02$ & $~~0.03 \pm 0.06$ & $~~0.00 \pm 0.06$ \\
68988 & $~~0.44 \pm 0.06$ & $~~0.38 \pm 0.03$ & $~~0.36 \pm 0.06$ & $~~0.37 \pm 0.06$ & $~~ 0.37 \pm 0.07$ \\
82943 & $~~0.30 \pm 0.06$ & $~~0.24 \pm 0.02$ & $~~0.22 \pm 0.02$ & $~~0.21 \pm 0.04$ & $~~0.22 \pm 0.04$ \\
83443 & $~~0.56 \pm 0.10$ & $~~0.45 \pm 0.05$ & $~~0.43 \pm 0.10$ & ---  & $~~0.43 \pm 0.10$ \\
95128 & $~~0.11 \pm 0.07$ & $~~0.05 \pm 0.02$ & $~~0.05 \pm 0.02$ & $~~0.07 \pm 0.05$ & $~~0.06 \pm 0.05$ \\
106252 & $~~0.04 \pm 0.02$ & $ -0.02 \pm 0.02$ & $ -0.10 \pm 0.02$ & $~~0.01 \pm 0.05$ & $ -0.04 \pm 0.04$ \\
108147 & $~~0.30 \pm 0.02$ & $~~0.22 \pm 0.02$ & $~~0.27 \pm 0.03$ & $~~0.28 \pm 0.07$ & $~~0.28 \pm 0.07$ \\
114783 & $~~0.25 \pm 0.11$ & $~~0.15 \pm 0.04$ & $~~0.27 \pm 0.08$ & $~~0.08 \pm 0.02$ & $~~0.17 \pm 0.07$ \\
117176 & $~~0.06 \pm 0.06$ & $ -0.03 \pm 0.02$ & $~~0.02 \pm 0.02$ & $~~0.04 \pm 0.06$ & $~~0.03 \pm 0.06$ \\
121504 & $~~0.14 \pm 0.05$ & $~~0.13 \pm 0.02$ & $~~0.14 \pm 0.05$ & $~~0.02 \pm 0.04$ & $~~0.08 \pm 0.06$ \\
136118 & $ -0.09 \pm 0.03$ & $~~0.00 \pm 0.02$ & $~~0.00 \pm 0.03$ & $ -0.12 \pm  0.02$ & $ -0.06 \pm 0.03$ \\
141937 & $~~0.16 \pm 0.04$ & $~~0.13 \pm 0.02$ & $~~0.12 \pm 0.03$ & $~~0.14 \pm 0.06$ & $~~0.13 \pm 0.06$ \\
160691 & $~~0.39 \pm 0.10$ & $~~0.33 \pm 0.04$ & $~~0.31 \pm 0.07$ & --- & $~~0.31 \pm 0.07$ \\
168746 & $~~0.12 \pm 0.04$ & $~~0.02 \pm 0.02$ & $~~0.11 \pm 0.02$ & $~~0.10 \pm 0.03$ & $~~0.10 \pm 0.03$ \\
169830 & $~~0.10 \pm 0.10$  & $~~0.19 \pm 0.02$ & $~~0.17 \pm 0.02$ & ---  & $~~0.17 \pm 0.02$ \\
190228 & $ -0.27 \pm 0.03$ & $ -0.27 \pm 0.03$ & $ -0.17 \pm 0.03$ & $ -0.29 \pm 0.06$ & $ -0.23 \pm 0.06$ \\
195019 & $ -0.23 \pm 0.25$ & $~~0.02 \pm 0.02$ & $ -0.02 \pm 0.02$ & $~~0.02 \pm 0.04$ & $~~0.00 \pm 0.04$ \\
202206 & $~~0.40 \pm 0.03$ & $~~0.31 \pm 0.03$ & $~~0.32 \pm 0.05$ & $~~0.27 \pm 0.05$ & $~~0.29 \pm 0.06$ \\
213240 & $~~0.33 \pm 0.10$ & $~~0.21 \pm 0.02$ & $~~0.29 \pm 0.02$ & --- & $~~0.29 \pm 0.02$ \\
\hline
\end{tabular}
\end{table*}

\begin{table*}
\centering
\caption{[X/H] values for Cr, Cu and Mn.}

\label{xmm}
\begin{tabular}{lccccc}
\hline
Star & Cr & Cr & Cr & Cu & Mn \\
(HD) & (EW) & (synth) & (avg) & & \\
\hline
4203 & $~~0.30 \pm 0.08$ & $~~0.23 \pm 0.07$ & $~~0.26 \pm 0.05$ & $~~0.48 \pm 0.06$ & $~~0.54 \pm 0.05$ \\
4208 & $ -0.36 \pm 0.06$ & $ -0.28 \pm 0.15$ & $ -0.34 \pm 0.06$ & $ -0.10 \pm 0.15$ & $ -0.19 \pm  0.06$ \\
6434 & $ -0.61 \pm  0.10$ & $ -0.55 \pm 0.07$ & $ -0.57 \pm 0.06$ & $ -0.49 \pm 0.05$ & $ -0.61 \pm  0.08$ \\
8574 & $ -0.08 \pm 0.06$ & $-0.18 \pm 0.07$ & $ -0.13 \pm 0.05$ & $~~0.01 \pm 0.05$ & $ -0.01 \pm  0.05$ \\
16141 & $~~0.11 \pm 0.05$ & $~~0.01 \pm 0.08$ & $~~0.06 \pm  0.04$ & $~~0.19 \pm 0.04$ & $ ~~0.24 \pm 0.06$ \\
19994 & $~~0.07 \pm 0.06$ & $-0.01 \pm 0.08$ & $~~0.03 \pm 0.05$ & $~~0.30 \pm 0.05$ & $ ~~0.31 \pm 0.05$ \\
22049 & $ -0.14 \pm 0.10$ & $-0.20 \pm 0.08$ & $ -0.18 \pm 0.06$ & $ -0.18 \pm 0.06$ & $ ~~0.16 \pm  0.06$ \\
27442 & ---  & ---  & ---  & ---  & $~~1.08 \pm 0.10$ \\
28185 & $~~0.15 \pm 0.05$ & $~~0.14 \pm 0.08$ & $~~0.15 \pm 0.04$ & $~~0.38 \pm 0.06$ & $~~0.52 \pm 0.07$ \\
33636 & $ -0.15 \pm 0.05$ & $-0.11 \pm 0.06$ & $ -0.13 \pm 0.04$ & $~~0.02 \pm 0.06$ & $ -0.01 \pm  0.04$ \\
37124 & $ -0.45 \pm 0.06$ & $ -0.37 \pm 0.05$ & $ -0.41 \pm 0.04$ & $ -0.21 \pm 0.04$ & $ -0.32 \pm  0.05$ \\
46375 & $~~0.16 \pm 0.06$ & $~~0.13 \pm 0.07$ & $~~0.15 \pm 0.04$ & $~~0.44 \pm 0.06$ & $~~0.61 \pm 0.06$ \\
50554 & $ -0.09 \pm 0.05$ & $-0.02 \pm 0.07$ & $-0.06 \pm 0.04$ & $~~0.08 \pm 0.05$ & $ ~~0.09 \pm 0.08$ \\
68988 & $~~0.29 \pm 0.06$ & $~~0.19 \pm 0.08$ & $~~0.24 \pm 0.05$ & $~~0.45 \pm 0.05$ & $~~0.61 \pm 0.05$ \\
82943 & $~~0.17 \pm 0.05$ & $~~0.26 \pm 0.05$ & $~~0.22 \pm 0.04$ & $~~0.40 \pm 0.05$ & $~~0.44 \pm 0.07$ \\
83443 & ---  & ---  & --- & --- & $~~0.88 \pm 0.06$ \\
95128 & $ -0.04 \pm 0.05$ & $-0.05 \pm 0.05$ & $-0.05 \pm 0.04$ & $~~0.16 \pm 0.04$ & $ ~~0.19 \pm 0.06$ \\
106252 & $ -0.16 \pm 0.05$ & $-0.10 \pm 0.08$ & $ -0.13 \pm 0.04$ & $~~0.06 \pm 0.05$ & $ ~~0.11 \pm 0.07$ \\
108147 & $~~0.14 \pm 0.10$ & --- & $~~0.14 \pm 0.10$ & --- & $~~0.37 \pm 0.07$ \\
114783 & $~~0.06 \pm 0.10$ & $-0.03 \pm 0.10$ & $~~0.01 \pm 0.08$ & $~~0.28 \pm 0.08$ & $~~0.39 \pm 0.10$ \\
117176 & $ -0.15 \pm 0.06$ & $-0.17 \pm 0.10$ & $ -0.16 \pm 0.05$ & $ -0.08 \pm 0.04$ & $ ~~0.15 \pm  0.06$ \\
121504 & $~~0.04 \pm 0.07$ & --- & $~~0.04 \pm 0.07$  & ---  & $~~0.39 \pm 0.06$ \\
136118 & $ -0.07 \pm 0.06$ & $-0.05 \pm 0.15$ & $-0.06 \pm 0.16$ & $~~0.11 \pm 0.15$ & $ ~~0.10 \pm 0.10$ \\
141937 & $~~0.10 \pm 0.06$ & $~~0.14 \pm 0.05$ & $~~0.12 \pm 0.04$ & $~~0.20 \pm 0.05$ & $ ~~0.30 \pm 0.03$ \\
160691 & --- & ---  & --- & --- & $~~0.62 \pm 0.07$ \\
168746 & $ -0.15 \pm 0.07$ & $ -0.21 \pm 0.08$ & $ -0.08 \pm 0.11$ & $~~0.01 \pm 0.04$ & $ ~~0.00 \pm  0.06$ \\
169830 & ---  & --- & --- & --- & $ ~~0.28 \pm 0.06$ \\
190228 & $ -0.35 \pm 0.06$ & $ -0.24 \pm 0.10$ & $ -0.30 \pm 0.06$ & $ -0.15 \pm 0.05$ & $ -0.19 \pm  0.07$ \\
195019 & $ -0.06 \pm 0.05$ & $-0.12 \pm 0.07$ & $-0.09 \pm 0.04$ & $ -0.01 \pm 0.07$ & $ ~~0.13 \pm 0.05$ \\
202206 & $~~0.27 \pm 0.06$ & $~~0.18 \pm 0.08$ & $~~0.23 \pm 0.05$ & $~~0.29 \pm 0.05$ & $~~0.56 \pm 0.05$ \\
213240 & --- & --- & --- & --- & $~~0.55 \pm 0.10$ \\
\hline
\end{tabular}
\end{table*}

\begin{table*}
\centering
\caption{[X/H] values fo Ni, Zn and Eu.}

\label{xmm}
\begin{tabular}{lccccc}
\hline
Star & Ni & Ni & Ni & Zn & Eu \\
(HD) & (EW) & (synth) & (avg) & & \\
\hline
4203 & $~~0.35 \pm 0.08$ & $~~0.29 \pm 0.05$ & $~~0.32 \pm 0.04$ & $~~0.56 \pm 0.08$ & $~~0.34 \pm 0.05$ \\
4208 & $ -0.23 \pm 0.07$ & $ -0.25 \pm 0.08$ & $ -0.24 \pm 0.05$ & --- & $~~0.03 \pm 0.04$ \\
6434 & $ -0.47 \pm 0.10$ & $ -0.54 \pm 0.10$ & $ -0.51 \pm 0.08$ & $ -0.50 \pm 0.10$ & $ -0.32 \pm   0.08$ \\
8574 & $~~0.02 \pm 0.06$ & $ -0.09 \pm 0.07$ & $ -0.03 \pm 0.05$ & --- & $ -0.11 \pm 0.06$ \\
16141 & $~~0.14 \pm 0.06$ & $~~0.03 \pm 0.06$ & $~~0.08 \pm 0.05$ & --- & $~~0.09 \pm 0.04$ \\
19994 & $~~0.00 \pm 0.07$ & $~~0.06 \pm 0.10$ & $~~0.03 \pm 0.06$ & $ -0.16 \pm 0.09$ & $~~0.00 \pm 0.06$\\
22049 & $ -0.14 \pm 0.09$ & $ -0.12 \pm 0.08$ & $ -0.13 \pm 0.07$ & --- & --- \\
27442 & $~~0.07 \pm 0.11$ & $~~0.65 \pm 0.10$ & $~~0.36 \pm 0.25$ & --- & --- \\
28185 & $~~0.17 \pm 0.06$ & $~~0.17 \pm 0.10$ & $~~0.17 \pm 0.05$ & $~~0.25 \pm 0.07$ & $~~0.05 \pm 0.04$ \\
33636 & $ -0.12 \pm 0.07$ & $ -0.12 \pm 0.06$ & $ -0.12 \pm 0.05$ & --- & $~~0.03 \pm 0.05$ \\
37124 & $ -0.30 \pm 0.07$ & $ -0.34 \pm 0.07$ & $ -0.32 \pm 0.05$ & --- & $ -0.23 \pm 0.07$ \\
46375 & $~~0.33 \pm 0.08$ & $~~0.24 \pm 0.08$ & $~~0.29 \pm 0.06$ & --- & $~~0.25 \pm 0.07$ \\
50554 & $ -0.04 \pm 0.06$ & $ -0.09 \pm 0.10$ & $ -0.06 \pm 0.06$ & --- & $ -0.02 \pm 0.05$ \\
68988 & $~~0.36 \pm 0.08$ & $~~0.38 \pm 0.07$ & $~~0.37 \pm 0.06$ & $~~0.38 \pm 0.09$ & $~~0.25 \pm 0.07$ \\
82943 & $~~0.26 \pm 0.07$ & $~~0.25 \pm 0.10$ & $~~0.25 \pm 0.07$ & --- & $~~0.22 \pm 0.05$ \\
83443 & $~~0.35 \pm 0.11$ & $~~0.45 \pm 0.09$ & $~~0.40 \pm 0.08$ & --- & $~~0.41 \pm 0.08$ \\
95128 & $~~0.02 \pm 0.06$ & $ -0.01 \pm 0.08$ & $~~0.00 \pm 0.05$ & $~~0.14 \pm 0.07$ & $ -0.11 \pm 0.04$ \\
106252 & $ -0.06 \pm 0.06$ & $ -0.11 \pm 0.08$ & $ -0.09 \pm 0.05$ & --- & $ -0.03 \pm 0.04$ \\
108147 & $~~0.35 \pm 0.09$ & $~~0.24 \pm 0.10$ & $~~0.30 \pm 0.07$ & --- & $~~0.48 \pm 0.05$ \\
114783 & $~~0.12 \pm 0.08$ & $ -0.04 \pm 0.10$ & $~~0.04 \pm 0.08$ & $ -0.43 \pm 0.10$ & $~~0.19 \pm 0.10$ \\
117176 & $ -0.11 \pm 0.08$ & $ -0.13 \pm 0.05$ & $ -0.12 \pm 0.05$ & $~~0.05 \pm 0.10$ & $~~0.02 \pm 0.08$ \\
121504 & $~~0.07 \pm 0.08$ & $~~0.16 \pm 0.08$ & $~~0.12 \pm 0.06$ & --- & $~~0.06 \pm 0.07$ \\
136118 & $ -0.17 \pm 0.08$ & $ -0.11 \pm 0.10$ & $ -0.14 \pm 0.07$ & $ -0.11 \pm 0.09$ & $~~0.01 \pm 0.06$ \\
141937 & $~~0.10 \pm 0.09$ & $~~0.13 \pm 0.05$ & $~~0.12 \pm 0.05$ & $~~0.34 \pm 0.09$ & $~~0.12 \pm 0.07$ \\
160691 & $~~0.23 \pm 0.09$ & $~~0.47 \pm 0.08$ & $~~0.35 \pm 0.08$ & --- & $~~0.51 \pm 0.06$ \\
168746 & $ -0.12 \pm 0.08$ & $ -0.08 \pm 0.06$ & $ -0.10 \pm 0.05$ & $ -0.23 \pm 0.09$ & $~~0.01 \pm 0.08$ \\
169830 & $ ~~0.14 \pm 0.08$ & $~~0.15 \pm 0.07$ & $~~0.15 \pm 0.05$ & --- & $~~0.15 \pm 0.06$ \\
190228 & $ -0.29 \pm 0.07$ & $ -0.33 \pm 0.07$ & $ -0.31 \pm 0.05$ & --- & $ -0.37 \pm 0.07$ \\
195019 & $ -0.11 \pm 0.06$ & $ -0.14 \pm 0.10$ & $ -0.12 \pm 0.06$ & --- & $ -0.15 \pm 0.04$ \\
202206 & $~~0.34 \pm 0.08$ & $~~0.23 \pm 0.10$ & $~~0.28 \pm 0.07$ & $~~0.22 \pm 0.09$ & $~~0.29 \pm 0.08$ \\
213240 & $~~0.22 \pm 0.08$ & $~~0.22 \pm 0.08$ & $~~0.22 \pm 0.06$ & --- & $~~0.08 \pm 0.08$ \\
\hline
\end{tabular}
\end{table*}

\begin{table*}
\centering
\begin{minipage}{70mm}
\caption{Calibration constants for equation 1 for elements other than C and O.}
\label{xmm}
\begin{tabular}{lccc}
\hline
Element & & [el/H] calibration constants\footnote{The abbreviations  are:
B -- \citet{bond06}; 
G -- \citet{gil06};
GG -- \citet{gg00}, \citet{gg01} and the present work;
LH -- \citet{luck06}.}: &\\
 & & A$_{\rm 0}$ &\\
 & & A$_{\rm 1} \times 10^{\rm 4}$&\\
 & & A$_{\rm 2}$ &\\
 & LH - G & GG - G & B - G\\
\hline
Na & $+1.72 \pm 0.19$ & $+0.80 \pm 0.17$ & $+2.03 \pm 0.36$\\
      & $-2.96 \pm 0.32$ & $-1.47 \pm 0.29$ & $-3.57 \pm 0.63$\\
      & $+0.18 \pm 0.04$ & $+0.07 \pm 0.05$ & $-0.24 \pm 0.06$\\
Mg & $+0.66 \pm 0.27$ & $-0.09 \pm 0.01$ & --\\
      & $-1.15 \pm 0.45$ & -- & --\\
      & $+0.29 \pm 0.07$ & -- & --\\
Al & $+0.69 \pm 0.15$ & $+0.04 \pm 0.02$ & $-0.30 \pm 0.01$\\
      & $-1.32 \pm 0.26$ & $-0.30 \pm 0.29$ & --\\
      & $+0.14 \pm 0.05$ & $+0.11 \pm 0.04$ & $-0.06 \pm 0.05$\\
Si & $+0.70 \pm 0.12$ & $+0.04 \pm 0.01$ & $+0.79 \pm 0.30$\\
      & $-1.15 \pm 0.20$ & -- & $-1.33 \pm 0.52$\\
      & $+0.15 \pm 0.03$ & $+0.07 \pm 0.04$ & $-0.13 \pm 0.05$\\
Ca & $+1.30 \pm 0.20$ & $+0.73 \pm 0.18$ & $+0.02 \pm 0.01$\\
      & $-2.06 \pm 0.34$ & $-1.09 \pm 0.32$ & --\\
      & $+0.31 \pm 0.06$ & $+0.17 \pm 0.07$ & $-0.16 \pm 0.06$\\
Sc & $-0.04 \pm 0.01$ & $+0.23 \pm 0.23$ & --\\
      & -- & $-0.45 \pm 0.39$ & --\\
      & $+0.25 \pm 0.05$ & $+0.25 \pm 0.07$ & --\\
Ti & $+0.70 \pm 0.16$ & $-0.02 \pm 0.17$ & $-1.02 \pm 0.81$\\
      & $-1.24 \pm 0.26$ & -- & $+1.6 \pm 1.4$\\
      & $+0.35 \pm 0.04$ & $+0.20 \pm 0.04$ & $+0.29 \pm 0.17$\\
Fe & $+0.71 \pm 0.13$ & $+0.22 \pm 0.09$ & $-0.12 \pm 0.02$\\
      & $-1.24 \pm 0.22$ & $-0.39 \pm 0.15$ & --\\
      & $+0.18 \pm 0.03$ & $+0.05 \pm 0.03$ & $-0.28 \pm 0.09$\\
Ni & -- & $+0.00 \pm 0.01$ & $-0.09 \pm 0.02$\\
      & -- & -- & --\\
      & -- & $+0.07 \pm 0.05$ & $-0.25 \pm 0.07$\\
\hline
\end{tabular}
\end{minipage}
\end{table*}

\begin{table*}
\centering
\begin{minipage}{70mm}
\caption{Calibration constants for equation 1 for C and O.}
\label{xmm}
\begin{tabular}{lcccc}
\hline
Element & & [el/H] calibration constants\footnote{The abbreviations for the studies are:
LH -- \citet{luck06};
E1 -- \citet{ecu04};
E2 -- \citet{ecu06b};
GG -- \citet{gg00}, \citet{gg01} and the present work;
T -- \citet{takhon05}.}: & &\\
 & & studies & &\\
 & & A$_{\rm 0}$ & &\\
 & & A$_{\rm 1} \times 10^{\rm 4}$ & &\\
 & & A$_{\rm 2}$ & &\\
\hline
C & LH - E1 & GG - E1 & T - E1 & B - E1\\
     & $-0.08 \pm 0.01$ & $+0.76 \pm 0.27$ & $+0.68 \pm 0.33$ & $-0.05 \pm 0.02$\\
     & -- & $-1.51 \pm 0.45$ & $-1.29 \pm 0.55$ & --\\
     & $+0.22 \pm 0.06$ & $+0.31 \pm 0.08$ & $+0.21 \pm 0.11$ & --\\
O  & GG - LH & T - LH & E2 - LH & --\\
     & $+0.92 \pm 0.41$ & $+0.60 \pm 0.45$ & $-0.05 \pm 0.03$ & --\\
     & $-1.65 \pm 0.69$ & $-1.06 \pm 0.77$ & -- & --\\
     & $+0.28 \pm 0.21$ & -- & $+0.53 \pm 0.18$ & --\\
\hline
\end{tabular}
\end{minipage}
\end{table*}

\begin{table*}
\centering
\caption{Combined abundance values for SWPs.}

\label{xmm}
\begin{tabular}{lccccccccccc}
\hline
Star & [Fe/H] & [C/H] & [O/H] & [Na/H] & [Mg/H] & [Al/H] & [Si/H] & [Ca/H] & [Sc/H] & [Ti/H] & [Ni/H] \\
(HD) & & & & & & $\sigma$,N & & & & & \\
\hline
142 & ~0.117 & -- & -- & ~0.218 & -- & -- & ~~0.144 & ~0.099 & -- & ~0.074 & ~0.113\\
 & 0.070, 2 & -- & -- & 0.070,2 & -- & -- & 0.061,2 & 0.125,2 & -- & 0.124,2 & 0.095,2\\
1237 & ~0.132 & -- & ~0.059 & ~0.005 & ~0.110 & ~0.109 & ~0.059 & ~0.015 & ~0.101 & ~0.111 & ~0.072\\
 & 0.047, 2 & -- & 0.076,2 & 0.051,2 & 0.045,2 & 0.051,2 & 0.043,2 & 0.074,2 & 0.081,2 & 0.060,2 & 0.051,2\\
2039 & ~0.323 & -- & -- & ~0.408 & -- & ~0.344 & ~0.315 & ~0.225 & -- & ~0.293 & ~0.331\\
 & 0.065, 2 & -- & -- & 0.065,2 & -- & 0.041,2 & 0.089,2 & 0.115,2 & -- & 0.104,2 & 0.065,2\\
4203 & ~0.388 & ~0.402 & ~0.147 & ~0.426 & ~0.425 & ~0.484 & ~0.375 & ~0.203 & ~0.462 & ~0.377 & ~0.360\\
 & 0.045, 2 & 0.075,2 & 0.151,2 & 0.061,2 & 0.082,2 & 0.047,2 & 0.051,2 & 0.073,2 & 0.086,2 & 0.075,2 & 0.063,2\\
4208 & -0.241 & ~0.009 & -- & -0.245 & -0.125 & -0.107 & -0.261 & -0.310 & -0.134 & -0.147 & -0.236\\
 & 0.035, 2 & 0.0149,2 & -- & 0.032,2 & 0.091,2 & 0.038,2 & 0.045,2 & 0.067,2 & 0.071,2 & 0.065,2 & 0.061,2\\
6434 & -0.520 & -0.145 & -- & -0.403 & -0.185 & -0.245 & -0.359 & -0.424 & -0.272 & -0.246 & -0.497\\
 & 0.075, 2 & 0.152,2 & -- & 0.051,2 & 0.086,2 & 0.061,2 & 0.045,2 & 0.087,2 & 0.102,2 & 0.083,2 & 0.071,2\\
8574 & ~0.029 & ~0.051 & ~0.093 & ~0.078 & ~0.102 & ~0.074 & ~0.007 & ~0.004 & ~0.099 & ~0.053 & -0.014\\
 & 0.053, 3 & 0.070,3 & 0.092,2 & 0.054,3 & 0.074,3 & 0.087,3 & 0.059,3 & 0.057,3 & 0.077,3 & 0.084,3 & 0.061,2\\
9826 & ~0.131 & ~0.220 & ~0.120 & ~0.225 & ~0.185 & -- & ~0.124 & ~0.081 & ~0.158 & ~0.131 & ~0.085\\
 & 0.067, 2 & 0.095,3 & 0.076,2 & 0.061,2 & 0.075,2 & -- & 0.045,2 & 0.085,2 & 0.108,2 & 0.076,2 & 0.095,2\\
10697 & ~0.144 & ~0.137 & ~0.099 & ~0.142 & ~0.205 & ~0.200 & ~0.090 & ~0.007 & ~0.190 & ~0.123 & ~0.101\\
 & 0.035, 2 & 0.065,3 & 0.074,2 & 0.030,2 & 0.051,2 & 0.025,2 & 0.022,2 & 0.054,2 & 0.070,2 & 0.035,2 & 0.061,2\\
12661 & ~0.346 & ~0.295 & -- & ~0.393 & ~0.425 & ~0.472 & ~0.291 & ~0.128 & ~0.381 & ~0.293 & ~0.314\\
 & 0.038, 2 & 0.055,2 & -- & 0.057,2 & 0.040,2 & 0.030,2 & 0.047,2 & 0.094,2 & 0.067,2 & 0.041,2 & 0.075,2\\
16141 & ~0.150 & ~0.119 & ~0.096 & ~0.094 & ~0.167 & ~0.194 & ~0.076 & ~0.024 & ~0.163 & ~0.155 & ~0.083\\
 & 0.037, 3 & 0.072,4 & 0.082,4 & 0.062,3 & 0.083,3 & 0.057,3 & 0.042,3 & 0.062,3 & 0.072,3 & 0.048,3 & 0.050,2\\
17051 & ~0.195 & ~0.215 & ~0.155 & ~0.208 & ~0.175 & ~0.167 & ~0.158 & ~0.103 & ~0.231 & ~0.160 & ~0.142\\
 & 0.056, 3 & 0.076,2 & 0.081,2 & 0.038,3 & 0.055,2 & 0.036,3 & 0.051,3 & 0.100,3 & 0.063,2 & 0.070,3 & 0.054,3\\
19994 & ~0.215 & ~0.277 & ~0.147 & ~0.401 & ~0.241 & ~0.280 & ~0.214 & ~0.127 & ~0.250 & ~0.175 & ~0.150\\
 & 0.058, 3 & 0.088,4 & 0.078,4 & 0.051,3 & 0.102,3 & 0.070,3 & 0.059,3 & 0.068,3 & 0.045,3 & 0.071,3 & 0.071,2\\
22049 & -0.121 & ~0.009 & -0.049 & -0.212 & -0.010 & -0.069 & -0.147 & -0.194 & -0.155 & -0.050 & -0.185\\
 & 0.035, 2 & 0.158,2 & 0.071,3 & 0.045,2 & 0.076,2 & 0.054,2 & 0.045,2 & 0.108,2 & 0.092,2 & 0.075,2 & 0.065,2\\
23079 & -0.124 & -- & -- & -0.086 & -- & -0.054 & -0.109 & -0.128 & -- & -0.049 & -0.171\\
 & 0.065,2 & -- & -- & 0.067,2 & -- & 0.041,2 & 0.055,2 & 0.096,2 & -- & 0.085,2 & 0.061,2\\
23596 & ~0.274 & ~0.249 & ~0.066 & ~0.429 & ~0.228 & ~0.320 & ~0.237 & ~0.148 & ~0.293 & ~0.241 & --\\
 & 0.050,2 & 0.045,2 & 0.050,2 & 0.051,2 & 0.108,2 & 0.079,2 & 0.055,2 & 0.047,2 & 0.067,2 & 0.082,2 & --\\
27442 & ~0.373 & ~0.358 & ~0.165 & ~0.315 & -- & ~0.531 & ~0.363 & ~0.049 & ~0.372 & ~0.423 & ~0.348\\
 & 0.098,2 & 0.469,2 & 0.115,2 & 0.177,2 & -- & 0.075,2 & 0.136,2 & 0.127,2 & 0.127,2 & 0.206,2 & 0.190,2\\
28185 & ~0.223 & ~0.224 & ~0.087 & ~0.267 & ~0.190 & ~0.315 & ~0.137 & ~0.038 & ~0.311 & ~0.266 & ~0.205\\
 & 0.056,3 & 0.066,3 & 0.093,3 & 0.081,3 & 0.087,3 & 0.060,3 & 0.105,3 & 0.077,3 & 0.085,3 & 0.068,3 & 0.045,2\\
30177 & ~0.384 & -- & -- & ~0.398 & -- & ~0.558 & ~0.387 & ~0.174 & -- & ~0.330 & ~0.421\\
 & 0.070,2 & -- & -- & 0.076,2 & -- & 0.061,2 & 0.075,2 & 0.130,2 & -- & 0.110,2 & 0.089,2\\
33636 & -0.072 & ~0.037 & -0.035 & -0.081 & -0.045 & -0.013 & -0.099 & -0.131 & -0.018 & -0.007 & -0.121\\
 & 0.045,3 & 0.090,4 & 0.086,4 & 0.033,3 & 0.076,2 & 0.054,3 & 0.080,3 & 0.064,3 & 0.067,3 & 0.108,3 & 0.073,2\\
37124 & -0.385 & -0.069 & ~0.015 & -0.352 & -0.115 & -0.101 & -0.285 & -0.360 & -0.177 & -0.134 & -0.374\\
 & 0.037,3 & 0.093,3 & 0.045,2 & 0.041,3 & 0.072,2 & 0.057,3 & 0.053,3 & 0.078,3 & 0.082,3 & 0.076,3 & 0.050,2\\
38529 & ~0.363 & ~0.327 & ~0.314 & ~0.431 & ~0.405 & ~0.448 & ~0.334 & ~0.186 & ~0.387 & ~0.332 & --\\
 & 0.062,3 & 0.110,3 & 0.082,3 & 0.065,2 & 0.129,2 & 0.096,2 & 0.082,2 & 0.065,2 & 0.075,2 & 0.098,2 & --\\
39091 & ~0.130 & -- & -- & ~0.179 & -- & ~0.156 & ~0.113 & ~0.033 & -- & ~0.065 & ~0.129\\
 & 0.063,2 & -- & -- & 0.051,2 & -- & 0.060,2 & 0.035,2 & 0.075,2 & -- & 0.055,2 & 0.050,2\\
40979 & ~0.214 & ~0.083 & -- & ~0.313 & ~0.209 & -- & ~0.203 & ~0.132 & ~0.191 & ~0.201 & --\\
 & 0.050,2 & 0.163,2 & -- & 0.040,2 & 0.079,2 & -- & 0.055,2 & 0.057,2 & 0.095,2 & 0.082,2 & --\\
46375 & ~0.234 & ~0.268 & ~0.145 & ~0.186 & ~0.350 & ~0.326 & ~0.208 & ~0.021 & ~0.232 & ~0.284 & ~0.221\\
 & 0.047,2 & 0.145,2 & 0.070,2 & 0.071,2 & 0.076,2 & 0.060,2 & 0.055,2 & 0.089,2 & 0.083,2 & 0.095,2 & 0.065,2\\
50554 & ~0.015 & ~0.048 & ~0.078 & ~0.027 & ~0.029 & ~0.040 & -0.042 & -0.058 & ~0.031 & ~0.022 & -0.063\\
 & 0.035,3 & 0.083,4 & 0.074,4 & 0.051,3 & 0.115,3 & 0.076,3 & 0.047,3 & 0.055,3 & 0.071,3 & 0.065,3 & 0.051,2\\
52265 & ~0.230 & ~0.209 & ~0.146 & ~0.278 & ~0.215 & ~0.264 & ~0.194 & ~0.110 & ~0.228 & ~0.189 & ~0.208\\
 & 0.047,3 & 0.068,4 & 0.068,4 & 0.043,3 & 0.069,3 & 0.083,3 & 0.053,3 & 0.052,3 & 0.062,3 & 0.065,3 & 0.067,2\\
68988 & ~0.341 & ~0.302 & ~0.245 & ~0.441 & ~0.377 & ~0.410 & ~0.322 & ~0.167 & ~0.367 & ~0.309 & ~0.358\\
 & 0.054,3 & 0.057,3 & 0.051,2 & 0.047,3 & 0.081,3 & 0.062,3 & 0.084,3 & 0.095,3 & 0.104,3 & 0.101,3 & 0.071,2\\
\hline
\end{tabular}
\end{table*}

\begin{table*}
\centering
\contcaption{}

\label{xmm}
\begin{tabular}{lccccccccccc}
\hline
Star & [Fe/H] & [C/H] & [O/H] & [Na/H] & [Mg/H] & [Al/H] & [Si/H] & [Ca/H] & [Sc/H] & [Ti/H] & [Ni/H] \\
(HD) & & & & & & $\sigma$,N & & & & & \\
\hline
70642 & ~0.202 & -- & -- & ~0.277 & -- & -- & ~0.187 & ~0.045 & -- & ~0.368 & ~0.270\\
 & 0.051,2 & -- & -- & 0.095,2 & -- & -- & 0.083,2 & 0.065,2 & -- & 0.073,2 & 0.051,2\\
72659 & ~0.008 & -0.026 & ~0.054 & ~0.058 & ~0.094 & ~0.051 & ~0.016 & -0.061 & ~0.069 & ~0.069 & --\\
 & 0.051,2 & 0.040,2 & 0.045,2 & 0.063,2 & 0.076,2 & 0.065,2 & 0.075,2 & 0.070,2 & 0.070,2 & 0.096,2 & --\\
73526 & ~0.266 & -- & -- & ~0.210 & -- & ~0.419 & ~0.263 & ~0.153 & -- & ~0.337 & ~0.274\\
 & 0.065,2 & -- & -- & 0.051,2 & -- & 0.055,2 & 0.067,2 & 0.110,2 & -- & 0.090,2 & 0.089,2\\
74156 & ~0.142 & ~0.148 & ~0.193 & ~0.218 & ~0.143 & ~0.236 & ~0.123 & ~0.054 & ~0.189 & ~0.162 & --\\
 & 0.045,2 & 0.045,2 & 0.045,2 & 0.030,2 & 0.030,2 & 0.114,2 & 0.045,2 & 0.070,2 & 0.078,2 & 0.081,2 & --\\
75289 & ~0.276 & ~0.201 & -- & ~0.198 & ~0.230 & ~0.286 & ~0.236 & ~0.194 & ~0.266 & ~0.216 & ~0.221\\
 & 0.064,3 & 0.082,2 & -- & 0.051,3 & 0.070,2 & 0.041,2 & 0.051,3 & 0.087,3 & 0.105,2 & 0.065,3 & 0.072,3\\
75732 & ~0.370 & ~0.292 & ~0.192 & ~0.278 & -- & ~0.517 & ~0.306 & ~0.091 & ~0.375 & ~0.340 & ~0.361\\
 & 0.061,2 & 0.107,3 & 0.105,2 & 0.081,2 & -- & 0.051,2 & 0.051,2 & 0.122,2 & 0.112,2 & 0.106,2 & 0.095,2\\
76700 & ~0.329 & -- & -- & ~0.336 & -- & ~0.537 & ~0.329 & ~0.236 & -- & ~0.332 & ~0.317\\
 & 0.067,2 & -- & -- & 0.071,2 & -- & 0.040,2 & 0.067,2 & 0.105,2 & -- & 0.061,2 & 0.095,2\\
80606 & ~0.349 & ~0.297 & -- & ~0.359 & ~0.383 & ~0.460 & ~0.301 & ~0.121 & ~0.381 & ~0.342 & --\\
 & 0.073,2 & 0.106,2 & -- & 0.075,2 & 0.148,2 & 0.060,2 & 0.071,2 & 0.114,2 & 0.116,2 & 0.100,2 & --\\
82943 & ~0.261 & ~0.263 & ~0.146 & ~0.307 & ~0.219 & ~0.257 & ~0.212 & ~0.122 & ~0.242 & ~0.201 & ~0.252\\
 & 0.034,3 & 0.090,4 & 0.072,4 & 0.024,3 & 0.071,3 & 0.066,3 & 0.047,3 & 0.048,3 & 0.064,3 & 0.054,3 & 0.057,2\\
83443 & ~0.377 & ~0.361 & ~0.155 & ~0.454 & -- & ~0.628 & ~0.432 & ~0.149 & ~0.490 & ~0.383 & ~0.454\\
 & 0.055,3 & 0.108,2 & 0.094,2 & 0.065,3 & -- & 0.079,3 & 0.047,3 & 0.088,3 & 0.105,2 & 0.100,3 & 0.084,3\\
89744 & ~0.256 & ~0.250 & ~0.219 & ~0.384 & ~0.363 & ~0.312 & ~0.222 & ~0.170 & ~0.254 & ~0.234 & ~0.217\\
 & 0.053,3 & 0.079,4 & 0.083,3 & 0.073,3 & 0.040,2 & 0.132,3 & 0.051,3 & 0.068,3 & 0.077,3 & 0.070,3 & 0.076,2\\
92788 & ~0.314 & ~0.285 & ~0.208 & ~0.369 & ~0.410 & ~0.408 & ~0.289 & ~0.170 & ~0.357 & ~0.309 & --\\
 & 0.044,3 & 0.096,2 & 0.108,2 & 0.051,2 & 0.045,2 & 0.060,2 & 0.063,2 & 0.076,2 & 0.065,2 & 0.065,2 & --\\
95128 & ~0.058 & ~0.124 & ~0.091 & ~0.083 & ~0.115 & ~0.100 & ~0.008 & -0.064 & ~0.112 & ~0.076 & ~0.011\\
 & 0.025,2 & 0.074,3 & 0.051,3 & 0.035,2 & 0.076,2 & 0.054,2 & 0.025,2 & 0.063,2 & 0.075,2 & 0.055,2 & 0.055,2\\
106252 & -0.037 & ~0.065 & ~0.108 & -0.019 & ~0.062 & ~0.031 & -0.067 & -0.137 & -0.010 & -0.032 & -0.092\\
 & 0.041,3 & 0.059,3 & 0.052,3 & 0.031,3 & 0.103,3 & 0.033,3 & 0.039,3 & 0.042,3 & 0.061,3 & 0.059,3 & 0.045,2\\
108147 & ~0.219 & ~0.178 & ~0.123 & ~0.206 & ~0.210 & -- & ~0.147 & ~0.121 & ~0.235 & ~0.194 & ~0.200\\
 & 0.055,2 & 0.076,2 & 0.067,2 & 0.051,2 & 0.076,2 & -- & 0.038,2 & 0.070,2 & 0.065,2 & 0.070,2 & 0.065,2\\
108874 & ~0.235 & ~0.224 & ~0.105 & ~0.167 & ~0.322 & ~0.294 & ~0.157 & ~0.023 & ~0.258 & ~0.216 & --\\
 & 0.050,2 & 0.110,2 & 0.120,2 & 0.061,2 & 0.074,2 & 0.057,2 & 0.051,2 & 0.076,2 & 0.100,2 & 0.075,2 & --\\
114762 & -0.657 & -0.316 & -0.162 & -0.546 & -0.313 & -- & -0.516 & -0.557 & -0.485 & -0.428 & --\\
 & 0.051,2 & 0.060,2 & 0.058,2 & 0.047,2 & 0.107,2 & -- & 0.051,2 & 0.070,2 & 0.075,2 & 0.098,2 & --\\
114783 & ~0.114 & ~0.308 & ~0.046 & -0.010 & ~0.205 & ~0.177 & ~0.059 & -0.174 & ~0.150 & ~0.160 & ~0.044\\
 & 0.032,2 & 0.135,2 & 0.129,2 & 0.051,2 & 0.076,2 & 0.071,2 & 0.051,2 & 0.075,2 & 0.096,2 & 0.075,2 & 0.085,2\\
117176 & -0.052 & -0.022 & ~0.015 & -0.121 & ~0.049 & ~0.081 & -0.092 & -0.154 & ~0.034 & ~0.006 & -0.116\\
 & 0.044,3 & 0.072,4 & 0.059,4 & 0.048,3 & 0.135,3 & 0.064,3 & 0.054,3 & 0.048,3 & 0.067,3 & 0.061,3 & 0.050,2\\
120136 & ~0.332 & ~0.329 & ~0.300 & ~0.463 & -- & ~0.398 & ~0.333 & ~0.161 & ~0.284 & ~0.330 & ~0.217\\
 & 0.079,3 & 0.084,4 & 0.055,2 & 0.096,3 & -- & 0.100,2 & 0.073,3 & 0.104,3 & 0.112,3 & 0.108,3 & 0.149,2\\
121504 & ~0.144 & ~0.130 & ~0.128 & ~0.107 & ~0.120 & ~0.188 & ~0.095 & ~0.113 & ~0.166 & ~0.149 & ~0.117\\
 & 0.050,2 & 0.091,2 & 0.071,2 & 0.025,2 & 0.082,2 & 0.055,2 & 0.032,2 & 0.149,2 & 0.085,2 & 0.060,2 & 0.065,2\\
128311 & ~0.023 & -- & -0.073 & -0.218 & ~0.096 & ~0.077 & -0.036 & -0.201 & ~0.138 & ~0.050 & --\\
 & 0.081,2 & -- & 0.126,2 & 0.139,2 & 0.071,2 & 0.095,2 & 0.070,2 & 0.114,2 & 0.126,2 & 0.136,2 & --\\
130322 & ~0.033 & ~0.017 & -0.060 & -0.103 & ~0.077 & ~0.084 & -0.036 & -0.140 & ~0.075 & ~0.047 & -0.005\\
 & 0.045,3 & 0.086,2 & 0.063,2 & 0.067,3 & 0.067,3 & 0.044,3 & 0.045,3 & 0.093,3 & 0.067,3 & 0.074,3 & 0.070,2\\
134987 & ~0.304 & ~0.307 & ~0.236 & ~0.342 & ~0.365 & ~0.396 & ~0.256 & ~0.110 & ~0.360 & ~0.269 & ~0.311\\
 & 0.062,4 & 0.080,4 & 0.081,4 & 0.060,4 & 0.064,3 & 0.044,4 & 0.059,4 & 0.087,4 & 0.051,3 & 0.078,4 & 0.089,3\\
136118 & -0.010 & ~0.049 & ~0.112 & ~0.067 & ~0.037 & ~0.023 & -0.042 & -0.057 & -0.007 & ~0.014 & -0.120\\
 & 0.053,3 & 0.081,3 & 0.045,2 & 0.075,3 & 0.085,3 & 0.082,2 & 0.058,3 & 0.090,3 & 0.062,3 & 0.092,3 & 0.057,2\\
141937 & ~0.121 & ~0.183 & ~0.076 & ~0.094 & ~0.172 & ~0.164 & ~0.074 & -0.004 & ~0.126 & ~0.117 & ~0.092\\
 & 0.044,3 & 0.070,3 & 0.067,3 & 0.034,3 & 0.108,3 & 0.066,3 & 0.039,3 & 0.050,3 & 0.074,3 & 0.069,3 & 0.050,2\\
143761 & -0.200 & -0.071 & ~0.035 & -0.175 & -0.019 & -- & -0.167 & -0.232 & -0.060 & -0.076 & --\\
 & 0.040,2 & 0.079,3 & 0.097,3 & 0.054,2 & 0.087,2 & -- & 0.035,2 & 0.076,2 & 0.081,2 & 0.086,2 & --\\
145675 & ~0.452 & ~0.364 & ~0.281 & ~0.452 & ~0.521 & ~0.546 & ~0.379 & ~0.131 & ~0.452 & ~0.413 & ~0.462\\
 & 0.079,3 & 0.110,4 & 0.110,3 & 0.091,3 & 0.102,3 & 0.074,3 & 0.049,3 & 0.131,3 & 0.118,3 & 0.141,3 & 0.100,2\\
\hline
\end{tabular}
\end{table*}

\begin{table*}
\centering
\contcaption{}

\label{xmm}
\begin{tabular}{lccccccccccc}
\hline
Star & [Fe/H] & [C/H] & [O/H] & [Na/H] & [Mg/H] & [Al/H] & [Si/H] & [Ca/H] & [Sc/H] & [Ti/H] & [Ni/H] \\
(HD) & & & & & & $\sigma$,N & & & & & \\
\hline
160691 & ~0.311 & ~0.260 & -- & ~0.406 & -- & ~0.389 & ~0.278 & ~0.115 & ~0.319 & ~0.268 & ~0.313\\
 & 0.045,3 & 0.076,2 & -- & 0.061,3 & -- & 0.062,3 & 0.052,3 & 0.076,3 & 0.076,2 & 0.068,3 & 0.073,3\\
168443 & ~0.076 & ~0.176 & ~0.069 & ~0.042 & ~0.224 & ~0.243 & ~0.075 & -0.040 & ~0.220 & ~0.162 & ~0.091\\
 & 0.044,3 & 0.060,4 & 0.069,4 & 0.031,3 & 0.048,3 & 0.078,3 & 0.045,3 & 0.069,3 & 0.065,4 & 0.064,3 & 0.067,2\\
168746 & -0.055 & ~0.092 & ~0.019 & -0.052 & ~0.128 & ~0.196 & -0.026 & -0.116 & ~0.131 & ~0.090 & -0.106\\
 & 0.041,3 & 0.096,3 & 0.058,3 & 0.033,3 & 0.083,3 & 0.071,3 & 0.047,3 & 0.073,3 & 0.099,4 & 0.057,3 & 0.055,2\\
169830 & ~0.188 & ~0.203 & ~0.151 & ~0.319 & ~0.211 & ~0.206 & ~0.138 & ~0.078 & ~0.141 & ~0.128 & ~0.116\\
 & 0.051,3 & 0.074,4 & 0.060,3 & 0.054,3 & 0.121,2 & 0.082,3 & 0.039,3 & 0.064,3 & 0.079,3 & 0.074,3 & 0.050,2\\
177830 & ~0.350 & ~0.436 & ~0.375 & ~0.305 & ~0.506 & ~0.550 & ~0.353 & -0.005 & ~0.389 & ~0.372 & --\\
 & 0.096,3 & 0.172,2 & 0.070,2 & 0.167,2 & 0.082,3 & 0.051,2 & 0.134,2 & 0.118,2 & 0.147,2 & 0.159,2 & --\\
178911 & ~0.239 & ~0.241 & ~0.147 & ~0.147 & ~0.273 & ~0.271 & ~0.217 & ~0.033 & ~0.270 & ~0.272 & --\\
 & 0.079,2 & 0.073,2 & 0.085,2 & 0.061,2 & 0.070,2 & 0.074,2 & 0.071,2 & 0.100,2 & 0.073,2 & 0.086,2 & --\\
179949 & ~0.226 & ~0.242 & ~0.099 & ~0.281 & ~0.274 & -- & ~0.156 & ~0.120 & ~0.137 & ~0.162 & --\\
 & 0.055,2 & 0.095,3 & 0.126,3 & 0.086,2 & 0.089,2 & -- & 0.067,2 & 0.070,2 & 0.085,2 & 0.079,2 & --\\
186427 & ~0.077 & ~0.086 & ~0.001 & ~0.074 & ~0.168 & ~0.142 & ~0.036 & -0.062 & ~0.156 & ~0.104 & --\\
 & 0.040,2 & 0.063,3 & 0.067,3 & 0.025,2 & 0.079,2 & 0.060,2 & 0.055,2 & 0.074,2 & 0.080,2 & 0.070,2 & --\\
187123 & ~0.125 & ~0.124 & ~0.036 & ~0.131 & ~0.169 & ~0.177 & ~0.070 & ~0.001 & ~0.137 & ~0.107 & ~0.099\\
 & 0.041,3 & 0.051,3 & 0.055,2 & 0.035,3 & 0.103,3 & 0.068,3 & 0.039,3 & 0.057,3 & 0.055,3 & 0.062,3 & 0.060,2\\
190228 & -0.236 & -0.155 & -0.140 & -0.263 & -0.104 & -0.102 & -0.262 & -0.316 & -0.146 & -0.149 & -0.289\\
 & 0.041,3 & 0.099,3 & 0.089,2 & 0.050,3 & 0.096,3 & 0.083,3 & 0.041,3 & 0.061,3 & 0.059,3 & 0.061,3 & 0.050,2\\
192263 & -0.039 & ~0.191 & ~0.062 & -0.243 & ~0.049 & ~0.028 & -0.041 & -0.242 & ~0.098 & ~0.029 & -0.108\\
 & 0.067,3 & 0.092,2 & 0.119,3 & 0.117,2 & 0.065,3 & 0.050,3 & 0.073,3 & 0.103,3 & 0.122,3 & 0.108,3 & 0.085,2\\
195019 & ~0.059 & ~0.089 & ~0.073 & ~0.011 & ~0.100 & ~0.132 & ~0.027 & -0.043 & ~0.041 & ~0.071 & -0.041\\
 & 0.037,3 & 0.072,4 & 0.076,4 & 0.038,3 & 0.092,3 & 0.036,3 & 0.039,3 & 0.067,3 & 0.154,3 & 0.062,3 & 0.051,2\\
196050 & ~0.228 & -- & -- & ~0.333 & -- & -- & ~0.261 & ~0.156 & -- & ~0.200 & ~0.267\\
 & 0.067,2 & -- & -- & 0.057,2 & -- & -- & 0.055,2 & 0.100,2 & -- & 0.076,2 & 0.081,2\\
202206 & ~0.330 & ~0.183 & ~0.231 & ~0.331 & ~0.348 & ~0.376 & ~0.242 & ~0.142 & ~0.340 & ~0.266 & ~0.261\\
 & 0.056,3 & 0.080,3 & 0.067,2 & 0.040,3 & 0.107,3 & 0.070,3 & 0.050,3 & 0.060,3 & 0.056,3 & 0.074,3 & 0.061,2\\
209458 & ~0.038 & ~0.049 & ~0.128 & ~0.017 & ~0.083 & ~0.059 & -0.010 & ~0.015 & ~0.067 & ~0.049 & ~0.005\\
 & 0.042,3 & 0.060,2 & 0.065,3 & 0.031,3 & 0.045,3 & 0.088,3 & 0.048,3 & 0.087,3 & 0.064,3 & 0.061,3 & 0.061,2\\
210277 & ~0.192 & ~0.241 & ~0.133 & ~0.170 & ~0.318 & ~0.384 & ~0.164 & ~0.039 & ~0.213 & ~0.242 & ~0.160\\
 & 0.059,3 & 0.076,4 & 0.066,3 & 0.085,3 & 0.114,3 & 0.060,3 & 0.039,3 & 0.088,3 & 0.095,3 & 0.085,3 & 0.067,2\\
213240 & ~0.210 & ~0.153 & ~0.122 & ~0.217 & -- & ~0.195 & ~0.132 & ~0.014 & ~0.248 & ~0.259 & ~0.162\\
 & 0.069,3 & 0.083,2 & 0.076,2 & 0.051,3 & -- & 0.039,3 & 0.043,3 & 0.143,3 & 0.095,2 & 0.061,3 & 0.071,3\\
216435 & ~0.238 & -- & -- & ~0.336 & -- & ~0.305 & ~0.232 & ~0.159 & -- & ~0.185 & --\\
 & 0.061,2 & -- & -- & 0.081,2 & -- & 0.038,2 & 0.054,2 & 0.079,2 & -- & 0.055,2 & --\\
216437 & ~0.269 & -- & -- & ~0.329 & -- & ~0.403 & ~0.232 & ~0.167 & -- & ~0.257 & ~0.252\\
 & 0.057,2 & -- & -- & 0.065,2 & -- & 0.035,2 & 0.071,2 & 0.096,2 & -- & 0.063,2 & 0.079,2\\
217014 & ~0.209 & ~0.201 & ~0.113 & ~0.243 & ~0.280 & ~0.277 & ~0.165 & ~0.056 & ~0.249 & ~0.169 & ~0.184\\
 & 0.048,3 & 0.069,4 & 0.082,4 & 0.041,3 & 0.040,2 & 0.071,3 & 0.054,3 & 0.050,3 & 0.075,3 & 0.081,3 & 0.051,2\\
217107 & ~0.346 & ~0.265 & ~0.194 & ~0.321 & ~0.393 & ~0.393 & ~0.292 & ~0.143 & ~0.338 & ~0.303 & ~0.354\\
 & 0.048,3 & 0.090,4 & 0.087,4 & 0.051,3 & 0.064,3 & 0.051,3 & 0.048,3 & 0.076,3 & 0.121,3 & 0.068,3 & 0.065,2\\
222582 & ~0.038 & ~0.032 & ~0.050 & ~0.029 & ~0.085 & ~0.115 & -0.013 & -0.098 & ~0.082 & ~0.016 & -0.049\\
 & 0.044,3 & 0.050,3 & 0.068,3 & 0.054,2 & 0.080,3 & 0.124,3 & 0.051,3 & 0.071,3 & 0.065,3 & 0.068,3 & 0.095,2\\

\hline
\end{tabular}
\end{table*}

\section*{Acknowledgments}

We thank Joan Gonzalez for help with data entry and the anonymous reviewer for very helpful comments.

\bsp

\label{lastpage}


\begin{thebibliography}{}

\bibitem[\protect\citeauthoryear{Beirao et al.}{2005}]{bei05} Beirao P., Santos N. C., Israelian G., Mayor M., 2005, A\&A, 438, 251
\bibitem[\protect\citeauthoryear{Bond et al.}{2006}]{bond06} Bond J. C., Tinney C. G., Butler R. P., Jones H. R. A., Marcy G. W., Penny A. J., Carter B. D., 2006, MNRAS, 370, 163
\bibitem[\protect\citeauthoryear{Cunha et al.}{2002}]{cunha02} Cunha K., Smith V. V., Suntzeff N. B., Norris J. E., Da Costa G. S., Plez B., 2002, AJ, 124, 379
\bibitem[\protect\citeauthoryear{Ecuvillon et al.}{2004}]{ecu04} Ecuvillon A., Israelian G., Santos N. C., Mayor M., Villan V., Bihain G., 2004, A\&A, 426, 619
\bibitem[\protect\citeauthoryear{Ecuvillon et al.}{2006}]{ecu06a} Ecuvillon A., Israelian G., Santos N. C., Mayor M., Gilli G., 2006, A\&A, 449, 809
\bibitem[\protect\citeauthoryear{Ecuvillon et al.}{2006}]{ecu06b} Ecuvillon A., Israelian G., Santos N. C., Shchukina N. G., Mayor M., Rebolo R., 2006, A\&A, 445, 633
\bibitem[\protect\citeauthoryear{Fischer \& Valenti}{2005}]{fv05} Fischer D. A., Valenti J., 2005, ApJ, 622, 1102
\bibitem[\protect\citeauthoryear{Gilli et al.}{2006}]{gil06} Gilli G., Israelian G., Ecuvillon A., Santos N. C., Mayor M., 2006, A\&A, 449, 723
\bibitem[\protect\citeauthoryear{Gonzalez}{1998}]{gg98} Gonzalez G. 1998, A\&A, 334, 221
\bibitem[\protect\citeauthoryear{Gonzalez}{2006a}]{gg06} Gonzalez G. 2006, MNRAS, 367, L37
\bibitem[\protect\citeauthoryear{Gonzalez}{2006b}]{ggpasp06} Gonzalez G., 2006, PASP, 118, 1494
\bibitem[\protect\citeauthoryear{Gonzalez \& Laws}{2000}]{gg00} Gonzalez G., Laws C., 2000, AJ, 119, 390
\bibitem[\protect\citeauthoryear{Gonzalez et al.}{2001}]{gg01} Gonzalez G., Laws C., Tyagi S., Reddy B. E., 2001, AJ, 121, 432
\bibitem[\protect\citeauthoryear{Gonzalez \& Vanture}{1998}]{gv98} Gonzalez G., Vanture, A. D., 1998, A\&A, 339,L29
\bibitem[\protect\citeauthoryear{Huang et al.}{2005}]{huang} Huang C., Zhao G., Zhang H.-W.., Chen Y.-Q., 2005, Chin. J. A\&A, 5, 619
\bibitem[\protect\citeauthoryear{Ida \& Lin}{2005}]{idalin05} Ida S., Lin D. N. C., 2005, Prog. Theor. Phys., 158, 68
\bibitem[\protect\citeauthoryear{Kurucz}{1993}]{kur93} Kurucz R., 1993, ATLAS9 Stellar Atmospheres  Programs and 2 km s$^{-1}$ Grid CD-ROM Vol. 13, Smithsonian Astrophysical Observatory
\bibitem[\protect\citeauthoryear{Kurucz \& Bell}{1995}]{kur95} Kurucz R., Bell B., 1995, Kurucz CD-ROM No. 23, Smithsonian Astrophysical Observatory
\bibitem[\protect\citeauthoryear{Laws et al.}{2003}]{laws03} Laws C., Gonzalez G., Walker K. M., Tyagi S., Dodsworth J., Snider K., Suntzeff, N. B., 2003, AJ, 125, 2664
\bibitem[\protect\citeauthoryear{Luck \& Heiter}{2006}]{luck06} Luck R. E., Heiter U., 2006, AJ, 131, 3069
\bibitem[\protect\citeauthoryear{Prochaska \& McWilliam}{2000}]{proch00} Prochaska J. X., McWilliam A., 2000, ApJ, 537, L57
\bibitem[\protect\citeauthoryear{Robinson et al.}{2006}]{rob06} Robinson S. E., Laughlin G., Bodenheimer P., Fischer D., 2006, ApJ, 643, 484
\bibitem[\protect\citeauthoryear{Santos et al.}{2005}]{sant05} Santos N. C., Israelian G., Mayor M., Bento J. P., Almeida P. C., Sousa S. G., Ecuvillon A., 2005, A\&A, 437, 1127
\bibitem[\protect\citeauthoryear{Sneden}{1973}]{sneden73} Sneden C., 1973, Ph.D. thesis, University of Texas at Austin
\bibitem[\protect\citeauthoryear{Takeda}{2003}]{tak03} Takeda Y., 2004, A\&A, 402, 343
\bibitem[\protect\citeauthoryear{Takeda \& Honda}{2005}]{takhon05} Takeda Y., Honda S., 2005, PASJ, 57, 65
\bibitem[\protect\citeauthoryear{Valenti \& Fischer}{2005}]{vf05} Valenti J., Fischer D. A., 2005, ApJS, 159, 141

\end{thebibliography}
\end{document}